\documentclass[preprint,aps,superscriptaddress,showpacs]{revtex4}
\usepackage{amsmath,amssymb,amsfonts,float}
\usepackage[dvips]{graphicx,color}
\usepackage{times}
\usepackage{pstricks,pst-node}
 
\newcommand{\bolS}{\mathbf{S}}

\newcommand{\bolr}{\mathbf{r}}
\newcommand{\bolk}{\mathbf{k}}

\newcommand{\boln}{\mathbf{n}}

\newcommand{\be}{\text{e}}

\newsavebox{\dotdot}
\savebox{\dotdot}[3mm]{\shortstack{\circle*{0.8}\\ \\ \circle*{0.8}}}

\begin{document}
\title{Geometric phases and the magnetization process 
in quantum antiferromagnets}
\author{Akihiro Tanaka}
\affiliation{Computational Materials Science Center, National Institute for 
Materials Science, Tsukuba 305-0047, Japan}
\affiliation{WPI Center for Materials Nanoarchitectonics, National Institute for 
Materials Science, Tsukuba 305-0047, Japan}
\author{Keisuke Totsuka}
\affiliation{Yukawa Institute for Theoretical Physics, Kyoto University, Kitashirakawa
Oiwake-Cho, Kyoto 606-8502, Japan}
\author{Xiao Hu}
\affiliation{WPI Center for Materials Nanoarchitectonics, National Institute for 
Materials Science, Tsukuba 305-0047, Japan}
\begin{abstract}
The physics underlying the magnetization process 
of quantum antiferromagnets is revisited from the viewpoint of geometric phases. 
A continuum variant of the Lieb-Schultz-Mattis-type 
approach to the problem is put forth, where the commensurability condition 
of Oshikawa {\it et al} derives from a Berry connection formulation of the 
system's crystal momentum. 
We then go on to 
formulate an effective field theory which can deal with higher dimensional 
cases as well. We find that a topological term,  
whose principle function is to assign Berry phase factors 
to space-time vortex objects, 
ultimately controls the magnetic behavior of the system. 
We further show how our effective action maps into a ${\bf Z}_2$ gauge theory under 
certain conditions, which in turn allows for the 
occurrence of a fractionalized phase with topological order.     
\end{abstract}            
\pacs{75.10.Jm} 
\maketitle
\section{Introduction}
\label{sec:intro}
Antiferromagnets subjected to an external magnetic 
field have attracted considerable attention over the years,  
largely owing to 
interesting features which they share 
with a rather broad class of quantum many-body systems 
situated on a lattice. 
The series of quantum phase transitions 
encountered as the field strength 
is varied may be viewed as a 
spin-analogue of the superfluid-Mott insulator transitions observed in 
boson Hubbard models\cite{Subir's-book,Totsuka1,Momoi-Totsuka}, 
which model e.g. Josephson junction arrays\cite{Fisher-W-G-F} and 
optical lattices\cite{Lewenstein-06}. 
Magnetization plateaus 
emerging at simple fractions of the saturated magnetization 
are reminiscent of the quantum Hall effects\cite{OYA,Misguich}. 
Recent 
developments reveal 
the subject to be interconnected with an even richer variety 
of issues such as the Luttinger theorem\cite{Yamanaka}, 
electric polarization in solids\cite{Resta,Nakamura-Voit}, 
and fractionalization of quantum numbers 
in dimensions greater than one\cite{Bonesteel,Oshikawa-Senthil}. 

In this article we revisit 
this problem from yet another perspective, 
i.e. that of geometric phases\cite{Shapere-Wilczek,Berry}. 
The basis of our arguments rests  
only on rather general properties of Berry connections, 
topological terms and boson-vortex duality, much of which are valid 
in any dimension. We aim to shed new light 
on the above body of interrelated phenomena,  
with particular emphasis on providing a workable format with which 
to pursue further exotica in quantum spin liquids,  
anticipated to emerge along this line of study. 
\section{Outline of this paper}
\label{sec:outline}
In view of 
the fairly technical nature of the presentation to follow,  
we wish to highlight in this section its main contents, placing them in context with 
previous work.      

Section \ref{sec:LSM} is devoted to a geometrical reinterpretation of the 
well-known $d$=1 result of Oshikawa, Yamanaka 
and Affleck (OYA)\cite{OYA}, which yields a quantum-mechanical constraint 
on the possible values which the magnetization can assume at plateaus, 
i.e. in a spin-gapped regime. 
(In this article the notation $d$ is reserved to denote the spatial dimensionality.)   
The OYA work (reviewed in subsection \ref{subsec:LSM}) 
builds on arguments initiated by 
Lieb, Schulz and Mattis\cite{LSM,Affleck-Lieb} (LSM), 
wherein the central step consists of comparing the 
crystal momenta of the ground state and a candidate low-lying excited state, 
constructed from the ground state by applying a slow twist to the spins. 
(The resulting quantization rule agrees with that obtained 
by bosonization methods  
\cite{Totsuka}.)  
Meanwhile there is a simple geometric formulation of the crystal momentum 
of a {\it ferromagnetic} spin chain due to Haldane\cite{Haldane-86}, 
which incorporates 
the language of spin Berry phases.   
We show that this framework can be 
adapted to our problem involving antiferromagnets, 
provided the spin moments are partially polarized due to the magnetic field. 
Carrying out the LSM procedure in this geometric language 
then simply amounts to evaluating the difference between Berry phases 
associated with untwisted and twisted spin configurations. We find that this  
indeed leads immediately to the celebrated OYA rule.   

In section \ref{sec:dual-vortex} we show that   
a similar geometric structure underlies the 
magnetization properties of antiferromagnets in 
{\em arbitrary} spatial dimensions.    
Here there are no obvious substitutes for  
the LSM scheme or the bosonization technique (both of which are by design 
specific to $d$=1), and the generalization of the OYA result 
to spatial dimensions larger than unity\cite{Oshikawa2} 
is known to be a subtle problem. 
Our strategy is to build instead on the semiclassical picture of the 
previous subsection and to derive a low-energy effective theory for 
partially polarized antiferromagnets. With the magnitude of the 
spin polarization essentially fixed by the magnetic field, 
the long wavelength physics mainly involves the orientational fluctuation 
of the residual {\it planar} staggered moment, 
lying in the plane perpendicular to the field. 
Symmetry thus dictates that our effective action should 
be a variant of the quantum XY model. 
A careful derivation confirms this expectation; we find however 
that the imaginary-time action for this XY model  
also contains a purely imaginary topological term (proportional to the 
time-derivative of the phase variable) whose coefficient is 
given by $S-m$, with $S$ the spin quantum number 
and $m$ the magnetization per site. 
This perhaps is not surprising if one recalls that a  
complex-valued low energy action also arises 
out of the closely-related boson-Hubbard model 
away from commensurate filling factors\cite{Fisher-Lee2,Subir's-book,Igor}. 

The later half of section \ref{sec:dual-vortex}  
thus explores   
the consequence of complexifying our action by 
the addition of such a term to the effective XY model.  
We know from earlier work on the hydrodynamical properties 
of superfluids and superconductors (where actions  
of the same form appear) that the topological term crucially influences the quantum dynamics of 
{\it phase vortices} by subjecting them to 
a Magnus-type force
\cite{Haldane-Wu,Fisher,Ao-Thouless}.    
At the heart of this phenomenon is an Aharonov-Bohm (AB)-like 
quantum phase interference 
which renders each vortex to act like a charged particle 
immersed in a fictitious magnetic field 
whose strength is fixed by the topological term. 
(The Magnus force is then 
simply understood to be a pseudo-Lorentz force.) 
Since vortex events (point vortices in (1+1)d, vortex loops in (2+1)d, 
and vortex sheets in (3+1)d) 
are the most relevant disordering agents of an XY model, 
it is clearly important to see how such an interference effect manifests itself 
in the present problem. 
For this purpose we incorporate simple duality techniques which  
enables us to visualize the physics directly in the language of vortices.  
(As later explained, 
the approaches taken here as well as in section \ref{sec:Z2-gauge}
are intimately related to the recent work of Balents et al\cite{competing orders} 
who (in a context slightly different from ours) run a detailed projective symmetry group analysis 
to classify the possible ground states 
which the vortices of the XY model can assume.) 
We find that 
the pseudo-AB effect generates Berry phase factors associated with 
the vortex events, which in turn governs (given a specified valued of $S-m$)  
whether the condensation of these topological defects is allowed as a result of constructive interference,  
or alternatively,  
a destructive interference renders the vortex configurations irrelevant. 
(The former case will imply spin disordering and hence the formation of a magnetization plateau.)  
It is easy to recognize that the former can happen for  
$S-m\in{\mathbb{Z}}$, the condition under which {\it all} vortex Berry phases become trivial. 
Defects of arbitrary vorticity (most importantly singly quantized vortices) 
will then be able to condense. 
A more subtle situation arises when the value of $S-m$ is a rational number, 
in which case vortices must be {\it multiply quantized} in order to condense.  
With this physical picture (which links Berry phases to the magnetization behavior) in hand, 
it is also interesting to study the effect of 
Dzyaloshinskii-Moriya type perturbations which can modify  
the Berry phase factors. A brief discussion on this issue is given at the end of this section.           

In section V we look into a particularly 
important class of problems where the quantity $S-m$ is a half-odd integer, 
i.e. $S-m\in{\mathbb{Z}}+1/2$. The central observation 
to make here is that the system can dynamically acquire  
a $\mathbb{Z}_2$ gauge symmetry. We thus come into contact with 
the work of Sachdev and Park\cite{Sachdev-Park} who, in the course of their 
search for novel spin liquids in the {\em absence} of a magnetic field 
arrive at the same effective theory (for easy-plane antiferromagnets)  
- a lattice XY model coupled to an Ising gauge theory 
with a $\mathbb{Z}_{2}$-valued Berry phase term. 
This theory has a phase diagram which accommodates a fractionalized 
phase\cite{Sachdev-Park,Sachdev-Park2}, 
and we discuss how this arises within our magnetization problem. 
Interestingly, we find that this derivation generalizes to the case  
where $S-m$ is set at other simple rational numbers; for instance 
when $S-m\in{\mathbb{Z}}+1/3$, 
we are lead to a similar effective theory coupled to a $\mathbb{Z}_{3}$ 
gauge field with a $\mathbb{Z}_{3}$-valued Berry phase term. 
The results of this section thus illustrates the unexpectedly rich 
phase structure of square lattice antiferromagnets in a magnetic field, 
suggesting them to be a promising place to seek  
exotic spin liquid states.  

Appendix \ref{appendix on OYA} supplements section \ref{sec:LSM} while 
the details of the duality methods used in 
section \ref{sec:dual-vortex} can be found in Appendix \ref{appendix on duality}.

\section{Geometrical interpretation of the Lieb-Schultz-Mattis argument}
\label{sec:LSM}
\subsection{A brief summary of Lieb-Schultz-Mattis}
\label{subsec:LSM}
The LSM argument\cite{LSM,Affleck-Lieb} gives us 
an important insight into how low-lying excitations in one dimensional 
systems are constructed.   Let us consider a translationally invariant 
$d$=1 quantum system with a unique ground state $|\text{G.S.}\rangle$.  
The key idea of LSM is that the following {\em twisted} state 
\begin{equation}
{\hat U}_{\rm LSM}|\text{G.S.}\rangle \equiv 
e^{i\sum_{j=1}^{N}\frac{2\pi j}{N} {\hat S}_{j}^{z}} |\text{G.S.}\rangle 
\end{equation}
is made up of low-lying excited states 
(provided that the crystal
momentum satisfies a certain condition 
which will be fixed shortly), and should hence be a useful reference state 
for extracting information on the low energy spectrum of the system.
To evaluate the momentum shift caused 
by ${\hat U}_{\rm LSM}$, it is convenient to consider the phase
acquired in the process of executing 
a 1-site translation $\hat{T}$ 
($\hat{T}\bolS_{n}\hat{T}^{-1} = \bolS_{n-1}$).   
Since  
\begin{equation}
\hat{T}\,{\hat U}_{\rm LSM}\hat{T}^{-1} 
= e^{-\frac{2\pi i}{N}\sum_{j}(S-S_{j}^{z})}{\hat U}_{\rm LSM} \; ,
\end{equation}
one can readily see that ${\hat U}_{\rm LSM}$ shifts the crystal
momentum $P$ by 
\begin{equation}
\delta P_{\text{LSM}} = -\frac{2\pi}{N}\sum_{j=1}^{N}(S-S_{j}^{z}) 
= -2\pi (S-m) \mod 2\pi \; ,
\label{eqn:LSM-shift}
\end{equation}
where we have introduced the magnetization (density)
$m=\sum_{j}S^{z}_{j}/N$.  
The above expression implies that if $S-m \not \in \mathbb{Z}$ 
the twisted state ${\hat U}_{\rm LSM}|\text{G.S.}\rangle$ is orthogonal 
to the ground state $|\text{G.S.}\rangle$, that is, 
the twisted state consists of excited states of the {\em finite-size} 
system.  
Moreover, it is not difficult to show that 
for generic Hamiltonians with short-range interactions 
the energy of the twisted state decreases like
$1/N$ as the system size grows $N\nearrow \infty$. 
Thus we are lead to an interesting dichotomy; 
if $S-m \not \in \mathbb{Z}$, the system in the thermodynamic limit 
either has gapless excitations over the unique ground state or 
has several degenerate ground states corresponding to a spontaneous 
breaking of (translational) symmetry. 
(The latter possibility becomes 
relevant when  
$S-m=p/q$  (where integers $p$ and $q$ are coprime), 
as one easily  
sees by repeating the above twist operation $q$ times.) 
This, combined with a complimentary commensurability argument 
based on bosonization, 
yields the so-called {\em quantization
condition} of magnetization plateaus\cite{OYA}: 
$Q_{\text{G.S.}}(S-m)\in \mathbb{Z}$ ($Q_{\text{G.S.}}$: period 
of the {\em infinite-size} ground state).  

In a strict sense, the above LSM argument alone tells us nothing about 
the presence/absence of plateaus as low-lying excitations created by 
the LSM twist do not change the total magnetization $\sum_{j}S^{z}_{j}$.  
Moreover, although a similar conclusion has been reached\cite{Oshikawa2}
in higher dimensions as well, it is not totally clear why the same sort of 
quantization condition exists 
regardless of dimensionality 
while the LSM argument 
is valid {\em only} in one dimension. 
In an attempt to clarify such issues, and to set the stage for 
unearthing further exotic properties,     
we will now reformulate the problem employing the language of Berry phases. 
%
Namely we will adopt a geometric approach to evaluating the  
crystal momenta of magnetic systems\cite{Haldane-86}, 
and show that 
the momentum of eq.(\ref{eqn:LSM-shift}), which was central to the LSM 
argument re-emerges as {\it an additional  
Berry phase} which is generated upon application of the twist. 
An incarnation of this geometric effect will arise in the form 
of vortex Berry phases in the low energy field theory 
described in section \ref{sec:dual-vortex}, 
which can be constructed for arbitrary $d$. 
\subsection{Berry phase argument}
We begin by quickly recalling how a spin Berry phase typically arises. 
Consider a spin-$S$ object represented 
by a spin coherent state\cite{Radcliffe} 
$\vert {\bf n}(t)\rangle$, where the unit vector ${\bf n}(t)$ specifies 
(in a semiclassical sense) the spin's orientation. 
When the dynamics of the system is such that ${\bf n}(t)$ 
undergoes an adiabatic rotation, returning to its initial value 
at the end of the excursion, the wavefunction accumulates a net phase of 
$S\oint_{c}d{\bf n}(t)\cdot {\bf a}
({\bf n}(t))=S\omega[{\bf n}(t)]$. 
Here 
${\bf a}({\bf n}(t))$ is a Berry connection defined by 
$\langle{\bf n}(t)\vert {\bf n}(t)+\delta{\bf n}\rangle
=\exp\left[iS{\bf a}({\bf n}(t))\cdot\delta{\bf n}\right]$, 
$C$ is the loop on the unit sphere mapped out by the 
trajectory $\{{\bf n}(t)\}$,  
and $\omega[{\bf n}(t)]$ the 
solid angle enclosed by $C$.  

One can conceive of situations where a similar 
phase accumulation 
is induced 
along a {\it spatial} (as opposed to temporal) extent 
of a many-spin system by a gradual spatial change of ${\bf n}$. 
Such an example 
was elaborated by Haldane\cite{Haldane-86} in his study 
of the crystal momenta of a 
ferromagnetic spin chain. 
Here one is concerned with an instantaneous spin 
configuration written as a direct product 
of spin coherent states, 
\begin{equation}
|\{ {\bf n}_{j}  \}\rangle\equiv 
\bigotimes_{j=1}^{N}\vert {\bf n}_{j}\rangle \; .
\label{eqn:coherent-state}
\end{equation} 
As in the previous subsection, information on the crystal momentum 
$P$ is gained   
by inspecting how the generator of translation 
${\hat T}=e^{i{\hat P}a}$ ($a$: lattice constant) affects 
this configuration.  
However, we should keep in mind that 
the state (\ref{eqn:coherent-state}) 
is in general not an eigenstate of $\hat{T}$, 
and we will in this semiclassical approach 
be evaluating the expectation value 
$\langle \{ {\bf n}_{j}  \} \vert {\hat T}\vert  \{ {\bf n}_{j}
\}\rangle$ instead of the eigenvalue itself\cite{Haldane-86}. 
Since by definition 
$\langle \{ {\bf n}_{j}  \} \vert
{\hat T}
\vert  \{ {\bf n}_{j}  \} \rangle
=
\prod_{j=1}^{N}
\langle{\bf n}_{j}\vert{\bf n}_{j-1}\rangle$, (we assume a 
periodic boundary condition (PBC), where the $(N{+}1)$th site 
is 
identified with site 1) we find, in 
analogy to 
the familiar temporal Berry phase, 
that in the continuum limit 
\begin{equation}
\langle \{ {\bf n}(x)  \} \vert
{\hat T}
\vert  \{ {\bf n}(x)  \}  \rangle
=
e^{iS\omega[{\bf n}(x)]}.
\label{Haldane's formula}
\end{equation} 
The solid angle $\omega[{\bf n}(x)]$ is 
now associated with the loop 
traced out by the snapshot configuration 
$\{{\bf n}(x)\}$. 
Eq.(\ref{Haldane's formula}) suggests that the 
crystal momenta of a $d$=1 ferromagnet 
is a topological quantity which is best described as a 
quantal phase defined modulo 2$\pi$.   

As illustrated in the previous subsection, 
evaluating the crystal momenta carried by the low-energy states 
is the central step in a LSM-type scheme\cite{LSM}. 
It is therefore tempting to derive 
the counterpart to eq.(\ref{Haldane's formula}) 
for our present problem,  
since by introducing such a relation into the LSM program, 
we can expect to arrive at a 
geometrical picture underlying the magnetization 
properties of antiferromagnets. 
We show below that this is indeed possible. 
In doing so however, 
the follow points require clarification.    
(1) The ground state of a quantum antiferromagnet is 
generally complicated, and cannot be expressed in the form of 
a spin coherent state as in eq.(\ref{eqn:coherent-state}). 
However, one can show \cite{footnote-1} that  
if we take the large-$S$ limit,  
the quantum ground state will generally  approach 
a conventional 
collinear N\'{e}el state. 
This is the basis of our use of a 
coherent state ansatz in the following semiclassical treatment. 
(We will explore in appendix \ref{appendix on OYA} what will happen 
when we relax this ansatz; see the remark of the end of this section.) 
(2) It is also pertinent to observe that eq.(\ref{Haldane's formula}) 
is sensible only when the spin orientation ${\bf n}(x)$ 
is of a smoothly varying nature. Meanwhile antiferromagnets in an external field 
would generally have components which vary on the lattice scale.   
For the Berry phase framework to work, therefore, we must devise 
a way to represent our system in terms of a coherent state labeled by some 
slowly-varying field.   

\begin{figure}[htb]
\begin{center}
\includegraphics[width=0.65\linewidth]{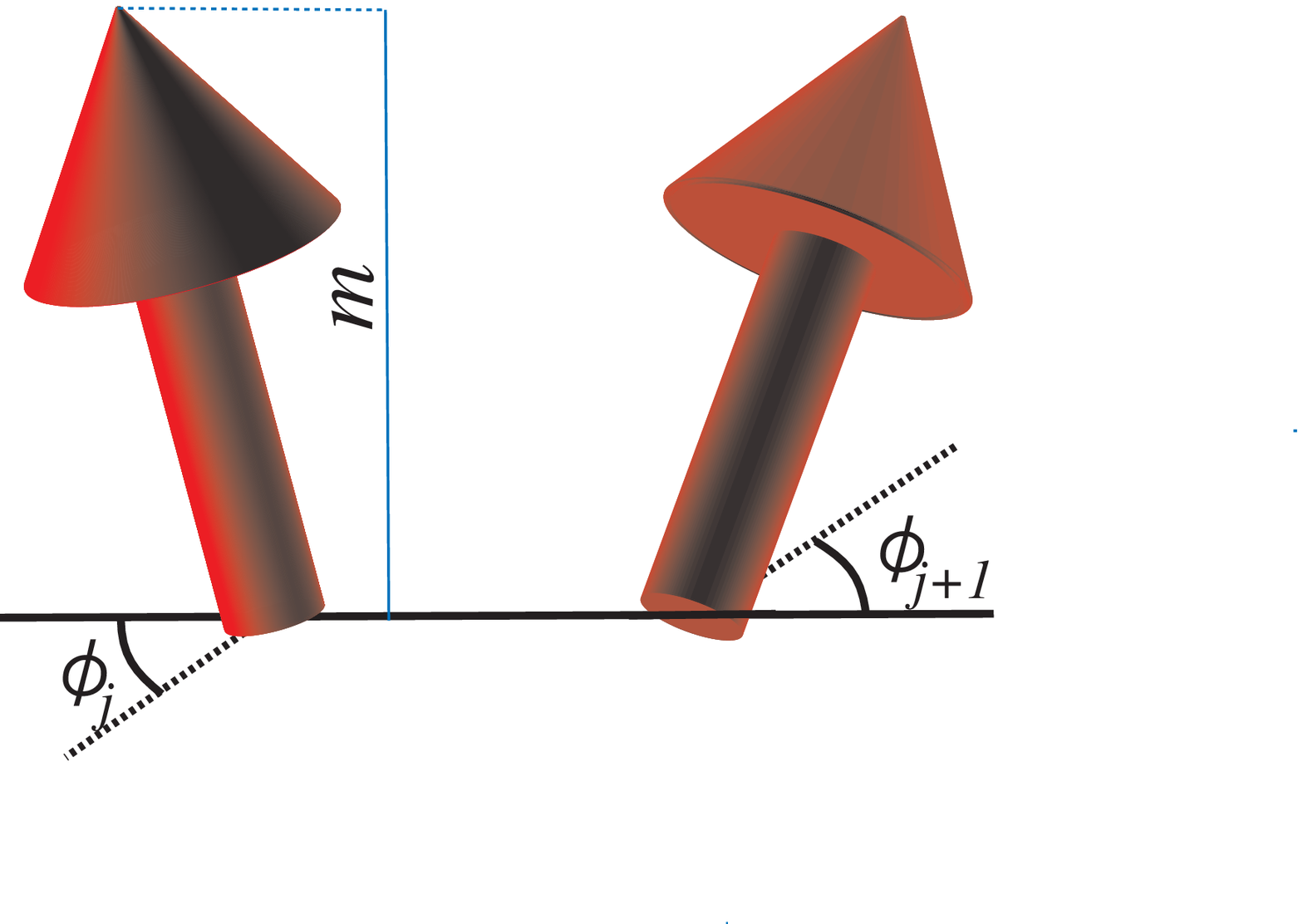}
\end{center}
\caption{
Schematic view of the slowly varying angular variables 
$\{\phi_{j}\}$, 
and the magnetization per site $m$. }
\label{angles figure}
\end{figure}

Let us start then from a canted configuration 
\begin{equation}
{\bf n}_{j}=((-1)^j \cos\phi_{j}\sin\theta_{j},
(-1)^j \sin\phi_{j}\sin\theta_{j},\cos\theta_{j}),
\end{equation}
where the unit vectors $\{{\bf n}_{j}\}$ represent the orientation of 
spins partially polarized 
by a magnetic field applied along the z-axis(see FIG.~\ref{angles figure}).  
We can identify the unit vector 
\begin{equation} 
{\bf N}_{j}\equiv(\cos\phi_{j}\sin\theta_{j},
\sin\phi_{j}\sin\theta_{j},\cos\theta_{j})
\end{equation}
as a slowly-varying (unstaggered) vector field 
for which the corresponding 
Berry connection 
\begin{equation}
S{\bf a}({\bf N(x)})
=
\langle\{{\bf N}(x)\}\vert 
(-i\nabla_{{\bf N}(x)})\vert \{{\bf N}(x)\}\rangle
\end{equation}
is a valid construction. 
This motivates us to introduce the unitary operator 
${\hat U}{\equiv} \exp\left[i\sum_{j=1}^{N}j\pi {\hat S}_{j}^{z}\right]$ 
which transforms 
$\vert \{ {\bf n}_{j} \}\rangle$ into $\vert \{ {\bf N}_{j} \}\rangle$, i.e. 
${\hat U}\vert \{ {\bf n}_{j} \}\rangle=\vert \{ {\bf N}_{j} \}\rangle$. 
Noting that 
${\hat T}{\hat U}{\hat T}^{-1}
=\exp\left[ i\pi\sum_{j=1}^{N}(S-{\hat S}_{j}^{z})\right]
{\hat U}$ 
(with PBC and $N$=even assumed), we find that 
\begin{eqnarray}
&\langle\{{\bf n}_{j}\}\vert
{\hat T}
\vert\{{\bf n}_{j}\}\rangle
=
\langle\{{\bf n}_{j}\}\vert
e^{ i\pi\sum_{j=1}^{N}(S-{\hat S}_{j}^{z})
}
{\hat U}^{-1}{\hat T}{\hat U}
\vert\{{\bf n}_{j}\}\rangle&\nonumber \\
&\simeq
e^{
i\pi\sum_{j=1}^{N}(S-S\cos\theta_{j})}
\langle\{{\bf N}_{j}\}\vert{\hat T}\vert\{{\bf N}_{j}\}\rangle,&
\end{eqnarray}
where the last line is a large-$S$ result and is a consequence of the 
minimal uncertainty property of the spin coherent state\cite{Takagi}. 
We may now incorporate Haldane's formula, eq.({\ref{Haldane's formula}}) 
to obtain 
\begin{equation}
\langle\{{\bf n}_{j}\}\vert
{\hat T}
\vert\{{\bf n}_{j}\}\rangle =e^{
i\pi\sum_{j=1}^{N}(S-S\cos\theta_{j})}
e^{iS\omega[{\bf N}(x)]}.
\label{main formula}
\end{equation}
Thus we have separated out the intrinsic part, where the Berry phase 
appears, from a momentum offset arising from the staggered nature of 
the ground state. 
(We pause to observe that in the 
{\it absence} of the magnetic field,  
a similar attempt 
to extract a Berry phase associated with 
smoothly varying components would run into difficulties,   
since that would involve, 
instead of a simple unitary transformation 
the simultaneous flipping of all three components of 
spins residing on every other site,   
i.e. time reversal operations.)   
The existence of this magnetization-dependent 
offset is also easily seen from the exact Bethe-Ansatz solution\cite{Yang-Yang} 
of the $S=1/2$ XXZ chain.   

The next task is to compare eq.(\ref{main formula}) 
with the corresponding expectation value 
for a LSM-twisted state. 
The latter state 
is defined by $\vert \{{\bf n}^{\rm LSM}_{j}\}\rangle
\equiv {\hat U}_{\rm LSM}\vert \{{\bf n}_{j}\}\rangle$, where 
${\hat U}_{\rm LSM}\equiv 
e^{i\sum_{j=1}^{N}\frac{2\pi j}{N} {\hat S}_{j}^{z}}$.
We note that $[{\hat U}, {\hat U}_{\rm LSM}]=0$. 
(Following OYA we are assuming that the interactions are sufficiently 
local and the Hamiltonian 
possesses a rotational symmetry around the z-axis.) 
The LSM-twisted counterpart of eq.(\ref{main formula}) is  
\begin{equation}
\langle\{{\bf n}_{j}^{\rm LSM}\}\vert
{\hat T}
\vert\{{\bf n}_{j}^{\rm LSM}\}\rangle
=e^{i\pi\sum_{j=1}^{N}(S-S\cos\theta_{j})}
e^{iS\omega[{\bf N}^{\rm LSM}(x)]},
\label{twisted formula}
\end{equation}
where ${\bf N}^{\rm LSM}(x)$ differs from ${\bf N}(x)$ by a shift 
of the azimuthal angle, i.e. 
$\phi_{\rm LSM}(x)\equiv \phi(x)-\frac{2\pi}{L}x$, $L=Na$. 
Notice that the twist leaves the offset portion of the momentum unaffected.   
The right hand side expressions of eqs.(\ref{main formula}) 
and (\ref{twisted formula}) 
will coincide if $S\omega[{\bf N}^{\rm LSM}(x)]=S\omega[{\bf N}(x)]+2\pi n$, 
where $n\in \mathbb{Z}$. Following the usual logic of LSM-type arguments we may 
interpret this as the semiclassical expression of the 
necessary condition for the occurrence of a spin gap, 
i.e. a magnetization plateau. We can make contact with the OYA theory by 
using the spherical coordinate representation for the solid angle, 
\begin{align}
S\omega[{\bf N}^{\rm LSM}(x)]
&=
S\int_{0}^{L}\! dx(1-\cos\theta(x))\partial_{x}
\phi_{\rm LSM}(x)
\notag \\
&=
S\omega[{\bf N}(x)] - \frac{2\pi S}{L}
\int_{0}^{L}\! dx(1-\cos\theta(x))
\notag \\
&=
S\omega[{\bf N}(x)] - 2\pi(S-m),
\label{solid angle}
\end{align}
where $m\equiv\frac{S}{L}\int_{0}^{L}dx\cos\theta$ is the 
magnetization density. This is the Berry-phase derivation of the LSM 
momentum shift (\ref{eqn:LSM-shift}).  
The aforementioned condition thus translates into  
the quantization rule $S-m\in\mathbb{Z}$. 
This argument is readily extended 
to the case where the unit cell consists of $r>1$ sites; 
the relevant quantity then will be the expectation value of ${\hat T}^{r}$, 
leading to the spin-gap condition 
$rS\omega[{\bf N}(x)]=rS\omega[{\bf N}^{\rm LSM}(x)]+2\pi n, 
n\in\mathbb{Z}$, which in turn yields 
the OYA quantization rule\cite{OYA,Totsuka} $r(S-m)\in\mathbb{Z}$. 

Finally a 
word on gauge independence is in order,  
as eq.(\ref{solid angle}) involves a particular gauge choice 
 for the monopole vector potential ${\bf a}({\bf N}(x))$ 
(the Dirac string goes through the south pole). 
One finds that relocating the string 
to the north 
pole merely shifts the 
crystal momentum by $4\pi S$,  
which 
is immaterial. 
Likewise, other gauge choices consistent with the 
spherical geometry of the 
target manifold will leave the results unaltered. 

As noted above, we have 
assumed on 
semiclassical grounds that it suffices to deal with    
ground states (and their twisted counterparts) which can be expressed as a 
spin coherent state.  
One may wonder though whether we can extend the present argument 
to more generic ground states which are {\it superpositions} 
of coherent states.  
We show that this is indeed possible in appendix \ref{appendix on OYA}.  
\section{Effective action and dual vortex theory}
\label{sec:dual-vortex}
The findings of the previous section tell us 
that 
for $d=1$, a simple semiclassical 
picture  
involving partially polarized 
spin moments is capable of 
accounting for the 
results of OYA, once we realize 
that certain quantum interference effects are at work. 
The latter are naturally described in terms of Berry phase concepts. 

This prompts us to seek a generalization of such geometric interpretations 
to arbitrary $d$.  
Since the LSM approach is now unavailable,  
we will  
shift gears and carefully work out an effective low energy theory valid 
in arbitrary $d$,   
invoking once more the semiclassical picture of canted spins. 
We will find that Berry phase effects are again present, 
now manifesting themselves in the form of 
quantum interference among vortex configurations. 
These interferences  can drastically alter the low energy properties of the system.  
(We will come full circle by later specializing to $d=1$, and identifying a 
common root that this effect shares with the  
LSM-type argument presented in section \ref{sec:LSM}.) We also discuss  
possible ways to perturb the system so that the Berry phases 
(and hence the magnetization properties) are modified.  
%
\subsection{Effective Action}
We now substantiate the foresaid by deriving a 
low energy effective action for an antiferromagnet coupled to 
an external magnetic field. 
The effective theory, summarized in eq.(\ref{quantum-XY-action}) below, 
is valid irrespective of the dimensionality.  

For the sake of concreteness, we will  
consider the following Hamiltonian: 
\begin{equation}
{\cal H} 
= J\sum_{\langle i,j \rangle}\bolS_{\bolr_{i}}{\cdot}
\bolS_{\bolr_{j}} + D\sum_{i}(S_{\bolr_{i}}^{z})^{2}
-H \sum_{i}S_{\bolr_{i}}^{z} 
\label{eqn:large-D-Ham}
\end{equation}
on a $d$-dimensional hypercubic lattice.  The first term corresponds to 
the usual Heisenberg Hamiltonian and the second to the single-ion
anisotropy. The external field $H$ is applied in the $z$-direction.  
If we regard the phase of $S^{+}$ and the local magnetization $S^{z}$ 
respectively as the Josephson angle $\phi$ and the boson density $n$, 
it is easy to see the similarity to the Josephson junction array:
\begin{equation}
{\cal H}_{\text{JJA}} = - \sum_{\langle i,j \rangle} 
J_{i,j}
\cos(\hat{\phi}_{\bolr_{i}}-\hat{\phi}_{\bolr_{j}})
+\frac{1}{2}\sum_{\langle i,j \rangle}V_{i,j}
\hat{n}_{\bolr_{i}}\hat{n}_{\bolr_{j}}
- \sum_{i} \mu_{\bolr_{i}}\hat{n}_{\bolr_{i}}  \; ,
\label{eqn:JJA-Hamiltonian}
\end{equation}
or, in more general terms, the boson Hubbard (BH) model.
In the large-$S$ limit of the model (\ref{eqn:large-D-Ham}), spins 
assume a canted (or,conical) configuration with the XY components aligned 
in an anti-parallel (antiferromagnetic) manner:
\begin{equation}
{\bf S}_{\bolr_{j}}=S \boln(\bolr_{j})
=\left(
\begin{array}{c}
{\sqrt{S^{2}-m_{j}^{2}}} \cos({\bf Q}{\cdot}{\bf r}_{j}) \\
{\sqrt{S^{2}-m_{j}^{2}}} \sin({\bf Q}{\cdot}{\bf r}_{j})\\
m_{j}
\end{array}
\right) \; ,
\label{unit vector}
\end{equation}
where $\mathbf{Q}=(\pi/a,\ldots,\pi/a)$.    
When either the field $H$ or the single-ion anisotropy $D$ is finite, 
a gap opens at $\bolk=\mathbf{Q}$ and the only gapless mode 
(a transverse spin wave) appears at $\bolk=\mathbf{0}$.  

Below we extract a low-energy 
effective theory from the model of eq.(\ref{eqn:large-D-Ham}), 
incorporating standard 
path-integral procedures\cite{Wen's-book,Auerbach's-book}. 
We illustrate this for the one-dimensional case. 
(Generalizing to higher dimensions is straightforward.)      
Guided by the classical solution, we parametrize the fluctuation 
around the above canted state 
as:
\begin{equation}
\begin{split}
& n^{\pm}(x) = (-1)^{r} \be^{\pm i\varphi(x)}\left(
\sin\theta^{(0)} + \delta\theta(x)\, \cos\theta^{(0)}
\right)  \\
& n^{z}(x) = \cos(\theta^{(0)}+\delta\theta(x)) 
\approx \cos\theta^{(0)} -\delta\theta(x)\, \sin\theta^{(0)} 
\; ,
\end{split}
\end{equation}
where $\theta^{(0)}= \cos^{-1}\left(\frac{H}{2S(2d J+D)} \right)$.   
The two equations above each correspond to modifying 
$\be^{i\mathbf{Q}{\cdot}\bolr_{j}}$ in (\ref{unit vector}) as 
$\be^{i\mathbf{Q}{\cdot}\bolr_{j}\pm i\varphi(x)}$, and expanding 
$m_{j}$ around the average $m= S \cos\theta^{(0)}$. 
An inspection of the classical Poisson bracket relation for the spin variables 
suggests that we identify 
the following as a pair of canonical variables  
($a$ denotes the lattice constant):
\begin{equation}
q(x) = \varphi(x) \; , \; p(x) = -S\,\sin\theta^{(0)} \delta\theta(x)
\equiv a \Pi(x)  \; ,
\end{equation}
which satisfy the equal-time commutator\cite{footnote-phi}
\begin{equation}
[\varphi(x)\, , \, \Pi(x^{\prime})] = i\,\delta(x-x^{\prime}). 
\end{equation}
From the expression 
\begin{equation}
S^{z}_{r} \approx S\,\cos\theta^{(0)} + a\,\Pi(x) + \cdots 
= m + a\Pi(x) + \cdots \; ,
\label{eqn:semiclassical-Sz}
\end{equation}
it is obvious that $\Pi$ describes the longitudinal fluctuations 
around the average magnetization $m$.  

\begin{widetext}
Casting these into path-integral form  
and retaining terms up to second order 
in $\varphi$ and $\Pi$ we obtain the action ${\cal S}_{\rm cl}+{\cal S}_{\rm BP}$ where:
\begin{eqnarray}
{\cal S}_{\text{cl}} 
&=& \int\!d\tau\int\!dx\, a(2J+D) \Pi^{2}(r) 
- \frac{1}{2}S^{2}
\left(1-\frac{H^{2}}{4S^{2}(D+2J)}\right)a \int\!d\tau\int\!dx\,
(\partial_{x}\varphi)^{2}
\nonumber \\
{\cal S}_{\text{BP}} &=&
i S\,(1-\cos\theta^{(0)})\sum_{r}\int\!d\tau\, \partial_{\tau}\varphi
-i \int\!d\tau\int\!dx\, \partial_{\tau}\varphi\, \Pi
\nonumber \\  
&=& i\frac{S-m}{a}\int\!dx d\tau\, \partial_{\tau}\varphi 
-i \int\!d\tau\int\!dx\, \partial_{\tau}\varphi\Pi .
\end{eqnarray}
The two contributations ${\cal S}_{\text{cl}}$ and 
${\cal S}_{\text{BP}}$ 
each come from the classical
Hamiltonian and the sum over the Berry phases of each spin.  
Although the first term of ${\cal S}_{\text{BP}}$ is a total
derivative, it cannot be dropped from the effective action 
for a reason which will become clear below.     
Following this intermediate step, we integrate out the $\Pi$-field 
to obtain the effective action for the angular field $\varphi$:  
\begin{equation}
{\cal S}_{\text{eff}} 
=
\int\!dxd\tau\, 
\Bigl\{
\frac{1}{2}\frac{\zeta}{v^{2}}(\partial_{\tau}\varphi)^{2} 
+ \frac{1}{2}\zeta (\partial_{x}\varphi)^{2}
\Bigr\}    
+i \frac{S-m}{a} \int\!dxd\tau\, \partial_{\tau}\varphi , 
\label{eqn:effective-action}
\end{equation}

where the spin stiffness $\zeta$ and the spinwave 
velocity $v$ are given by
\begin{equation}
 \zeta = a J S^{2} \left(
1-\frac{H^{2}}{4(D+2J)^{2}S^{2}} \right), 
{\mbox{\hspace*{3mm}}}
v = Ja\sqrt{\frac{4(D+2J)^{2}S^{2}-H^{2}}{2J(D+2J)}}.
\end{equation}
Additional interactions will merely renormalize the values of $\zeta$ and $v$.  
Note that the $\varphi$-field appearing in (\ref{eqn:effective-action}) 
is defined on a universal covering space of a circle and the last term 
counts the winding number of the space-time history. 
In section \ref{sec:Z2-gauge} 
we will show that it is also possible to  
arrive at this term by carefully summing over the spin Berry phase terms 
$iS\omega[{\bf n}({\rm r}_{j})]$ associated with each lattice site. 
\end{widetext}

In the above, we have assumed 
that in the transverse (XY, here) direction, 
the slowly-varying degree of freedom is a staggered component 
with wavevector ${\bf Q}$, i.e. possesses at least a short-range antiferromagnetic order. 
However, the argument goes exactly in the same manner for 
the spiral (helical) magnets as well; the external field kills two of the three 
Goldstone modes and the remaining one is described again by $\varphi$. 
Therefore, our effective action (\ref{eqn:effective-action}) 
is applicable equally well to non-frustrated- and frustrated cases. 

Our derivation of ${\cal S}_{\text{eff}}$ 
(\ref{eqn:effective-action}) 
for $d=1$ 
readily generalizes to any spatial dimension $d$, and leads to  
the following effective action:
\begin{align}
& {\cal S}_{\text{eff}}[\varphi(\tau,x)]
= {\cal S}_{\rm top}[\varphi(\tau,{\bf r})]
+{\cal S}_{\rm XY}[\varphi(\tau,{\bf r})],\nonumber\\
& {\cal S}_{\rm top}
\equiv i\int d\tau d^{d}{\bf r}\rho\partial_{\tau}\varphi\nonumber\\
& {\cal S}_{\rm XY} \equiv 
\int d\tau d^{d}{\bf r}
\left[\frac{K_{\tau}}{2}(\partial_{\tau}\varphi)^{2}
+\frac{K_{\perp}}{2}(\nabla \varphi)^{2} \right]
\label{quantum-XY-action}
\end{align}
where 
$\rho\equiv\frac{S-m}{a^{d}}$,  
$K_{\tau}=\zeta/v^{2}$ and 
$K_{\perp}=\zeta$.  
Eq.(\ref{quantum-XY-action}) 
bears the form of an XY model in $(d+1)$
dimensions supplemented with a topological term. 
An identical action 
was previously employed to describe the hydrodynamical 
properties of superfluids\cite{Haldane-Wu,Fisher,Ao-Thouless}. 
In this analogy, $\rho$ plays the role of a uniform offset value of the 
superfluid density, as one can read off from eq.(\ref{offsetted lagrangian}) 
of appendix \ref{appendix on duality}. 
There is also an apparent similarity to the low energy theory for the 
BH model at 
incommensurate filling factors\cite{Fisher-Lee2,Igor}, which is natural 
in view of the correspondence between eqs.(\ref{eqn:large-D-Ham}) and 
(\ref{eqn:JJA-Hamiltonian}); we will come back to this connection later.
As already mentioned, the topological term ${\cal S}_{\rm top}$ is a total derivative 
and as such will not affect the classical equation of motion. 
It does however 
have profound influences on low energy physics  
when the quantum effects of space-time vortices, 
to which we now turn, 
are properly accounted for. 
\subsection{Dual Vortex Theory}
\label{sec:vortex-ham}
The sensitivity of the topological term to vortices  
is already apparent from the simple observation that a 
phase winding (in imaginary time) 
of $2\pi$ at spatial position ${\bf r}_{j}$ will yield   
a nonvanishing contribution $\delta {\cal S}_{\rm top}=i2\pi\rho$.
To understand the consequence of such 
effects, it proves convenient to apply 
to eq.(\ref{quantum-XY-action}) a standard boson-vortex 
duality transformation\cite{Lee-Fisher} 
and recast the action in terms of 
vortex events: point-like space-time vortices (phase-slips) in (1+1)d, 
vortex loops in (2+1)d (world-lines of point-like vortex-antivortex pairs), 
and closed vortex surfaces in (3+1)d (world-sheets of vortex rings). 
For brevity we will generally resort to continuum notations 
while keeping track of the lattice origin of our model. 
Following steps well-accounted for in the literature\cite{Lee-Fisher,Zee}, 
we extract the following action 
(derivations are supplied in appendix \ref{appendix on duality}),
\begin{equation}
{\cal S}_{\rm vortex}={\cal S}_{\text{Coulomb}} 
+{\cal S}^{\text{vortex}}_{\text{BP}},
\end{equation}
where ${\cal S}_{\rm Coulomb}$ represents the ($d$+1) dimensional 
Coulombic interaction among space-time vortex objects 
mediated by the kernel 
\begin{equation}
\left(\frac{1}{-\partial^{2}}\right)
\equiv -\left(\frac{1}{2K_{\perp}}(\partial_{\tau})^{2}
+\frac{1}{2K_{\tau}}(\nabla)^{2}\right)^{-1},
\end{equation}
and ${\cal S}^{\rm vortex}_{\rm BP}$ is the Berry phase associated with 
vortex events.  
The latter inherits the 
information on the ``superfluid density'' 
$\rho\propto S-m$ 
contained in the original topological term and 
takes the following forms:
\begin{subequations} 
\begin{equation}
{\cal S}_{\text{BP}}^{\text{vortex}} = 
\begin{cases}
i\frac{2\pi(s-m)}{a}\sum_{j}q_{j}a^{(0)}_{j}  \quad & \text{(1+1)d} \\
i\frac{2\pi(s-m)}{a^{2}}\sum_{j,\mu}l_{j,\mu}a^{(0)}_{j,\mu}  
\quad & \text{(2+1)d} \\
i\frac{2\pi(s-m)}{a^{3}}\sum_{j,\mu\nu}
v_{j,\mu\nu}a^{(0)}_{j,\mu\nu}  \quad & \text{(3+1)d} \; ,
\end{cases}
\end{equation}
where $a^{(0)}$ is a solution to 
\begin{equation}
\begin{cases}
- \epsilon_{\mu\nu}\partial_{\nu}a^{(0)}_{j} = \delta_{\mu,\tau} 
\quad & \text{(1+1)d} \\
\epsilon_{\mu\nu\rho}\partial_{\nu}a^{(0)}_{j,\rho} = \delta_{\mu,\tau} 
\quad & \text{(2+1)d} \\
-\epsilon_{\mu\nu\rho\sigma}\partial_{\nu}a^{(0)}_{j,\rho\sigma} = \delta_{\mu,\tau} 
\quad & \text{(3+1)d} \; .
\end{cases}
\label{eqn:eq-of-offset}
\end{equation}
By using the explicit solution to eq.(\ref{eqn:eq-of-offset}), 
${\cal S}^{\rm vortex}_{\rm BP}$ can be written compactly as:
\begin{equation}
{\cal S}^{\rm vortex}_{\rm BP}
=
\begin{cases}
i\frac{2\pi (S-m)}{a}\sum_{j}q_{j}X_{j} \quad & \text{(1+1)d} \\
i\frac{2\pi (S-m)}{a^2}\sum_{j}q_{j}A^{xy}_{j} & \text{(2+1)d} \\
i\frac{2\pi (S-m)}{a^3}\sum_{j}q_{j}V^{xyz}_{j} & \text{(3+1)d} \; .
\end{cases}
\label{vortex BP factors}
\end{equation}
\end{subequations}
The summations on the right hand side are to be taken over 
all vortex events, and the $q_{j}$'s are their vorticities. 
$X_{j}$ denotes the spatial coordinate of the 
$j$-th (1+1)d vortex, 
$A^{xy}_{j}$ the area bounded by the 
projection onto the $xy$-plane of the $j$-th vortex loop 
$C_j$ in (2+1)d. 
$V^{xyz}_{j}$ is the volume of 
the j-th vortex surface in (3+1)d projected 
onto the xyz space (i.e. the net real-space volume 
occupied by a vortex-ring through its lifetime). 
\begin{figure}[htb]
\begin{center}
\includegraphics[width=0.8\linewidth]{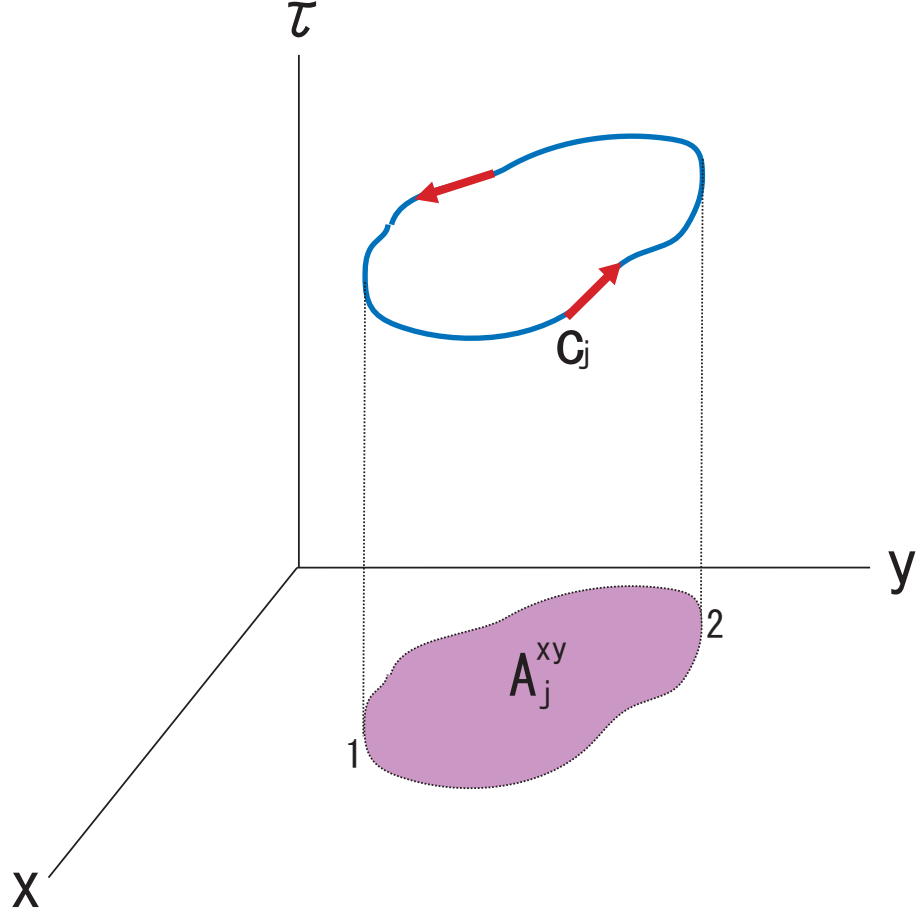}
\end{center}
\caption{
A vortex loop event $C_{j}$ in (2+1)d. 
Projected onto the xy plane, this corresponds to 
a vortex-antivortex pair created 
at point 1, which pair-annihilates at point 2. }
\label{area figure}
\end{figure}

The implications of the vortex Berry phase factors are clear. 
In (1+1)d, the profile of vortex events, 
when projected onto the spatial coordinate axis consists of points   
residing on dual sites. 
The spatial coordinate difference of any pair of vortex events 
(not necessarily occurring at equal times),  
$X_{i}-X_{j}$, are therefore 
integer multiples of $a$. 
Likewise, in (2+1)d and (3+1)d, $A^{xy}_{j}=a^{2}\times$integer 
and $V^{xyz}_{j}=a^{3}\times$integer, respectively. 
Hence when $S-m\notin\mathbb{Z}$, 
vortices will create contributions to the partition function 
that are weighted by oscillatory phase factors, 
generally leading to a destructive interference (unless events with 
vorticities of equal magnitudes and opposite signs are confined 
in pairs). 
(For the moment we are leaving aside the commensurability effects 
which set in when 
$S-m$ is a rational non-integer number.)
In other words vortices are, under this condition unable to  
contribute to the partition function. 
In (1+1)d, for instance, 
a Kosterlitz-Thouless transition into the plasma phase is prohibited 
quantum mechanically regardless of the 
value of the coupling constants.\cite{Wen's-book,Horovitz} 
When $S-m\in \mathbb{Z}$, in contrast, the phase factors trivialize and 
the partition function reduces to that 
of the ($d+1$) dimensional XY-model. 
Vortices are then able to proliferate and condense, 
if energetically favorable, and drive the system into a disordered state.


With the hindsight of the results just mentioned, we  
briefly reflect upon the LSM approach of the previous section 
and attach to it a simple physical picture. 
Given that the system in question can 
basically be regarded as a (1+1)d XY model subject to a periodic 
spatial boundary condition, 
we shall consider a one dimensional superfluid 
(with superfluid density $\rho_s$) 
in a ring geometry, a setup often employed to discuss the stability of persistent supercurrents.     
The LSM twist in this language is nothing but {\it a phase slip process} 
which causes the total phase difference along the ring to change 
by $\pm 2\pi$. It is well known\cite{Wen's-book,Khlebnikov} that such an event will 
generate a Galilei boost, i.e. an increment of the center-of-mass momentum of the 
superfluid by the amount $\Delta P\equiv
\int dx \rho\partial_{x}\varphi=\pm 2\pi\rho$. 
The latter leads to 
a full agreement with the findings of section \ref{sec:LSM} provided    
that we are allowed to equate $\rho_s$ 
with the coefficient of the topological term, $\rho=\frac{S-m}{a}$. 
As mentioned earlier, a short inspection of eq.(\ref{offsetted lagrangian}) 
in appendix \ref{appendix on duality}
confirms that the topological term indeed functions as a source term for 
the superfluid density, imposing precisely this value.    
In this way we find that the vortex Berry phases displayed in eq.(\ref{vortex BP factors}) 
and the momentum increment due to the LSM twist share a common origin\cite{Paramekanti}; 
the topological information encoded in the 
term ${\cal S}_{\rm top}$ of eq.(\ref{quantum-XY-action}). 
Within the present physical picture, the novelty of the 
condition $S-m\in \mathbb{Z}$ may be understood 
by observing   
that it allows for {\it the extra phase winding due to the phase slip 
to occur without the need for the superfluid to transfer excess momenta to phonons}. 
(If not for this condition, such transfer will be generically forbidden at zero temperature.)    
In other words, for this special case it is possible (if energetically favored) 
for the ground state to be a superposition of different circulation numbers 
$\frac{1}{2\pi}\oint dx{\partial_{x}\varphi}\in \mathbb{Z}$, 
i.e. a condensate of phase slips.

\begin{widetext}
While we have so far treated the vortices 
at a ``first-quantized" level, 
the Berry phase factors of 
eq.(\ref{vortex BP factors}) are readily incorporated into 
second-quantized vortex-field theories\cite{Banks,Peskin,Igor}, 
which provide a framework 
more amenable to detailed analysis. 
Referring to the later half of appendix \ref{appendix on duality} for details, 
we display here their main structures: 
\begin{subequations}
\begin{align}
&{\cal S}^{\text{(1+1)d}}_{\rm dual}[\widetilde{\varphi}]
= {\cal S}_{\rm kin}[\widetilde{\varphi}]
+y\int d^{2}r\cos[2\pi(S-m)x+2\pi\widetilde{\varphi}]& \\
&{\cal S}^{\text{(2+1)d}}_{\rm dual}[{\widetilde b}_{\mu},\chi]
= {\cal S}_{\rm kin}[{\widetilde b}_{\mu}] 
+\sum_{\mu}\frac{1}{2\lambda}\int d^{3}r
\cos[2\pi(\Delta_{\mu}\chi-{\widetilde b}_{\mu}-\delta_{\mu y}(S-m)x)] &
\\
&{\cal S}^{\text{(3+1)d}}_{\rm dual}[{\widetilde b}_{\mu\nu},c_{\mu}]
= {\cal S}_{\rm kin}[{\widetilde b}_{\mu\nu}] 
+\sum_{\mu<\nu}\frac{1}{2\lambda}\int d^{4}r
\cos[2\pi(\Delta_{\mu}c_{\nu} -\Delta_{\nu}c_{\mu})  
-2\pi({\widetilde b}_{\mu\nu}
+\delta_{\mu y}\delta_{\nu z}(S-m)x)]& \; .
\end{align}
\label{vortex-field-theory}
\end{subequations}
\end{widetext}
The kinetic terms denoted by ${\cal S}_{\rm kin}$ are quadratic; 
${\cal S}_{\rm kin}^{(1+1)d}$ is a gaussian theory, while 
the higher-dimensional counterparts are Maxwellian terms 
associated with the dual gauge 1- and 2-forms, 
${\widetilde b}_{\mu}$ and ${\widetilde b}_{\mu\nu}$. 
In principle, the action can be supplemented with additional terms 
consistent with the symmetry requirements of the problem.  
As explained in appendix \ref{appendix on duality}, 
$\partial_{x}{\widetilde \varphi}$ ($d$=1) and 
the gauge curvatures $\epsilon_{\tau\mu\nu}\partial_{\mu}{\widetilde b}_{\nu}$ ($d$=2), 
$\epsilon_{\tau\mu\nu\lambda}\partial_{\mu}{\widetilde b}_{\nu\lambda}$ ($d$=3) are directly related to the 
local magnetization (or in the superfluid analogy, to the superfluid density).
The 
fields $\chi$ and $c_{\mu}$ are introduced into the theory to implement 
the continuity of the vortex current (2+1)d/vortex-loop current 
(3+1)d\cite{Banks,Peskin,Igor}. 
For each 
case the Berry phase effect enters as a spatial 
modulation $2\pi(S-m)x$ within the cosine term, again causing 
oscillatory behavior unless the 
forementioned condition is met. 
Similar expressions (including higher harmonics terms) have been discussed in 
(1+1)d using bosonization methods. 
\cite{Totsuka,Totsuka1
}

It is worth noting that for (2+1)d, the above expression 
is identical in form to the free energy of a (classical) 3d lattice 
superconductor, subjected to an external magnetic field of strength 
$2\pi(S-m)$ (expressed in the Landau gauge). 
The problem at hand therefore reduces in this case to the energetics 
of a system of Abrikosov vortices immersed in a dual lattice superconductor;  
if these vortices manage to destruct superfluidity in this dual theory, 
it is tantamount to a massless superfluid state of the original XY model.   
If on the other hand superfluidity remains intact in the dual theory, 
the original theory is disordered, i.e., is massive.  
This observation enables us to 
make contact with a related argument due to Lee and
Shankar\cite{Lee-Shankar} (LS), 
who also employ a duality technique (which is somewhat different from ours, 
and in particular do not explicitly emphasize the relevance of Berry phases) 
to map quantum 2d lattice systems into an ensemble of Abrikosov vortices. 
Following LS, it is natural to expect that when $S-m$ is an integer 
(or more generally, is a rational number-a case we had not explicitly 
addressed up to now) 
the ground state is a periodic vortex lattice 
commensurate with and pinned by the underlying lattice structure 
of our spin system.  
For irrational $S-m$, they will form a floating lattice 
and destroy the superfluidity of the dual superconductor. 

The case where $S-m$ is a non-integer rational number, 
should however be 
treated with special caution, since     
subtle quantum effects may lead to novel physics    
which the naive superconductor energetics can miss\cite{Oshikawa2}.  
To illustrate what can happen, we go back to the vortex Berry phases of 
eq.(\ref{vortex BP factors}), and put $S-m=p/q$, $q\ge 2$. 
Singly quantized vortices/antivortices ($q_{j}=\pm 1$) are clearly frustrated; 
they will suffer the accumulation of an 
AB-like phase of $\frac{2\pi p}{q}$ upon encircling a dual plaquette. 
This leads to the destructive interference that we had been describing 
earlier in this section. 
Vortices whose winding numbers are integer multiples of $q$, 
on the other hand are free of such effects.    
This leads us to expect that a highly unusual phase can appear, 
in which only vortices with vorticity $\pm q$ 
are able to condense; such a tendency can be stabilized by the $q$-th 
higher harmonics of the 
cosine term of eq.(\ref{vortex-field-theory}), 
which is the leading non-oscillating term in the harmonic expansion. 
(Needless to say this commensurability effect becomes weak when $q$ is 
too large.) 
Since a condensate of 
higher-winding number vortices can sustain fractional 
excitations\cite{Balents1,Balents2}, 
this simple argument points to the possibility of yielding 
a fractionalized phase in antiferromagnets 
in an applied magnetic field. Actually an even richer variety of 
phases are possible in this 
commensurate case, and can be investigated in its full generality 
by adopting methods initiated 
by Lannert et al \cite{Lannert}, 
and further developed by other 
authors\cite{Sachdev-Park,Tesanovic,competing orders,Kim-san}. 
It 
is readily seen 
that there are actually $q$ degenerate low energy modes within 
the magnetic Brillouin zone for 
frustrated vortices hopping on a square (dual) lattice under the influence of 
a $\frac{2\pi p}{q}$-flux piercing each (dual) plaquette. 
The natural procedure would thus be to construct a Ginzburg-Landau-like 
action involving $q$ species of vortex fields, 
and to study the condensation of various composites of such objects. 
We will not work out the details of this approach here as  
they can be found in the literature\cite{competing orders} 
(albeit in different contexts).  
In section \ref{sec:Z2-gauge} we will instead focus on 
the important case where $S-m$ 
is half-integer-valued, and illustrate the nature of possible 
fractionalized phases 
which can emerge. 
  
\subsection{Correspondence with the Boson Hubbard Model}
It is instructive to understand how the above conforms 
with properties of the well-known chemical potential vs. 
tunneling parameter phase diagram of the 
BH model\cite{Fisher-W-G-F}.  
The most generic feature of the BH phase diagram are lobes of 
Mott insulator phases lined up along the chemical potential 
($\mu$) axis in the weak tunneling region.  
The state within each lobe, being incompressible, 
is characterized by a constant integer-valued boson density $N$. 
As evident from the foregoing 
(see also the discussion of Sachdev\cite{Subir's-book}), there is a  
set of correspondences here with the present problem which may be 
summarized as: 
$\mu\leftrightarrow H$, $N\leftrightarrow S-m$, 
Mott insulator$\leftrightarrow$spin gapped phase, 
superfluid phase$\leftrightarrow$gapless phase. 
The BH counterpart to our problem of 
determining the gapfulness/gaplessness of the antiferromagnet 
at fixed $S-m$ thus amounts to examining, at a fixed value of $N$, 
whether it is possibility for a superfluid to directly enter 
a Mott insulator phase  upon the approach to the weak tunneling regime. 
Here we would like to place this 
analogy on firmer grounds by recalling how Berry phases 
affect the critical properties of the BH model.    

It is known\cite{Fisher-W-G-F} that when 
$N\in \mathbb{Z}$ the 
constant-$N$ contour in the superfluid phase merges with the  
Mott insulator phase (with the same $N$-value) at the tip of the lobe. 
Meanwhile the contours for noninteger-$N$ will disappear 
into the region in between two adjacent Mott insulator lobes; 
the fate of these contours (especially those representing 
commensurate fillings) is a delicate matter\cite{van Otterlo} 
and will ultimately depend on the short-range physics. 
The former situation, in which a superfluid-to-insulator transition 
is generically possible, 
is therefore reminiscent of the $S-m\in\mathbb{Z}$ case of 
the antiferromagnets.  
The reason behind this similarity can be traced to the 
behavior of the topological term present in the 
low energy action; namely, 
both systems correspond to 
cases where this term ceases to be effective. 
We have already discussed in length the irrelevance of Berry phase effects 
for the $S-m\in\mathbb{Z}$ antiferromagnets. 
Meanwhile the topological term for the effective action of the BH model is 
strongly constrained by a U(1) gauge symmetry which the original Hamiltonian 
possessed\cite{Subir's-book,Igor}. This constraint can be used to show 
that the coefficient of this term 
vanishes identically right at the lobe tip. Hence the quantum phase 
transition which takes place along the 
integer-$N$ contour at this point is in the universality of 
the $(d+1)$ dimensional XY model, 
which is the same universality class relevant to the antiferromagnets 
with $S-m\in\mathbb{Z}$.      

\subsection{Effects of antisymmetric interactions}
In closing this section we briefly discuss a possible 
way of modifying the vortex Berry phases we have discussed above. 
Consider, in the superfluid analogy, allowing a bulk supercurrent 
${\bf J}$ to flow through a $d$-dimensional sample; 
this corresponds to perturbing the spin system with  
a Dzyaloshinskii-Moriya (DM) type interaction 
\begin{equation}
{\cal H}_{\text{DM}}=
\sum_{\bf r}\sum_{a=1}^{d}D_{a}{\hat z}{\cdot}({\bf S}_{\bf r}
{\times}{\bf S}_{{\bf r}+{\hat e}_{a}}) \; .
\end{equation} 
It is easy to see that this perturbation 
induces a non-zero spin current 
(which translates into a supercurrent), 
and can be accounted for in the effective theory 
of eq.(\ref{quantum-XY-action}) by simply adding 
the term $i J_{a}\nabla_{a}\varphi$, where $J_{a}\sim D_{a}$ is 
the induced supercurrent.  
This addition will obviously effect the vortex Berry phases. 
For instance, while the (2+1)d expression in eq.(\ref{vortex BP
factors}) 
basically counts the number of bosons residing within the shaded 
region of Fig.2, 
we must now correct for the migration due to the flow. 
(Physically, this will cause a change in the quantum dynamics of each vortex. 
Besides the Magnus force (perpendicular to the velocity of the vortex) 
coming from the original topological term, 
vortices will suffer an additional ``Lorentz force" 
which is perpendicular to the supercurrent.)  
This shifts the Berry phase term by 
\begin{equation}
\Delta{\cal S}^{\text{(2+1)d}}_{\text{BP}}
=i (2\pi) \sum_{j\in\text{loops}}q_{j}(J_{x}A^{y\tau}_{j}+J_{y}A^{\tau x}_{j}) \; ,
\end{equation} 
where $A^{y\tau}_{j}$ and $A^{\tau x}_{j}$ are each the area of 
the projected image of $C_{j}$ onto the $y\tau$- and $\tau x$ planes.  
A similar consideration leads to the following Berry-phase shift in
(1+1)-dimensions:
\begin{equation}
\Delta{\cal S}^{\text{(1+1)d}}_{\text{BP}} =
i (2\pi) J_{x}\sum_{j}q_{j}\tau_{j} \; .
\end{equation} 

Now let us consider 
how the non-zero supercurrent $J_{a}$ 
(or, non-zero DM interactions) affects the magnetization process for the
simplest (1+1)-dimensional case.  
In accordance with eq.(\ref{vortex-field-theory}), the low-energy physics 
in the vicinity of $S-m \in \mathbb{Z}$ may be described 
by the following sine-Gordon model:
\begin{equation}
{\cal S}^{\text{(1+1)d}}_{\rm dual}[\widetilde{\varphi}]
= {\cal S}_{\rm kin}[\widetilde{\varphi}]
+y\int d^{2}r\cos[2\pi (\delta m\, x + J_{x}\tau) + 2\pi\widetilde{\varphi}]
\end{equation}
where $\delta m$ measures a small deviation from $S-m\in \mathbb{Z}$. 

By a straightforward extension of the methods used 
in Refs.\onlinecite{Jose-K-K-N,Horovitz}, we can carry out 
a renormalization-group analysis.   
According to the 1-loop renormalization-group equations,  
there are two different length scales: (i) a length scale 
$\xi(\delta m,J_{x})\equiv 1/\sqrt{(\delta m)^{2}+(J_{x})^{2}}$ set both by 
$\delta m$ and by $J_{x}$ (or, $D$) and (ii) another set by the lowest 
particle (i.e. $\delta S^{z}\neq 0$) excitation gap 
$\Delta_{\text{sol}}$ in the absence of $\delta m$ and $J_{x}$ (if it is finite).  
When $\Delta_{\text{sol}}=0$, nothing competes with $\delta m$ or $J_{x}$ 
and the system is in the gapless (Tomonaga-Luttinger liquid) phase. 
When $\Delta_{\text{sol}} \neq 0$, on the other hand, 
we have two different possibilities depending on the ratio between 
$\xi(\delta m,J_{x})$ and $\Delta^{-1}_{\text{sol}}$; 
if $\xi(\delta m,J_{x}) < \Delta^{-1}_{\text{sol}}$, renormalization 
stops well before the system starts feeling the effect of the pinning 
potential and the system flows into the same gapless phase as above.  
If $\xi(\delta m,J_{x}) > \Delta^{-1}_{\text{sol}}$, 
the system is renormalized into another phase.  
The nature of this phase 
may be analyzed by mapping the system to a system of (nearly) free 
fermions\cite{Schulz}; 
we find that the gap responsible for the plateau is robust against $J_{x}$, i.e. the DM 
interaction $D$.    

\section{Nature of plateau phases -- Z$_{2}$ gauge theory}
\label{sec:Z2-gauge}
In the previous section, we had seen that for 
$S{-}m\in \mathbb{Z}$ vortices do not suffer from interference effects 
due to the Berry phase (\ref{vortex BP factors}).  
As a consequence, the usual $(d{+}1)$-dimensional XY 
transition\cite{Fisher-W-G-F,Shenoy} 
separates the gapless XY-ordered (superfluid) phase 
at large-$K_{x,\tau}$ from the small-$K$ plateau (i.e. insulating) phase 
where vortices proliferate.   
For generic incommensurate values of $S-m$, on the other hand, 
destructive interference among different 
configurations of topological (vortex) excitations should lead to 
quite different small-$K$ behaviors.  
However, it is not easy to capture the fine structures which may emerge  
as we approach the limit $K\rightarrow 0$. 
For this reason, we utilize a slightly different formulation 
of the problem\cite{Sachdev-Park} 
more suited to study the strong coupling phases,  
and attempt to see more closely how the topological excitations 
destroy the gapless superfluid phases.  

The derivation presented below, leading to our effective theories rely 
heavily on methods described in ref.\onlinecite{Sachdev-Park}. 
We are lead however to interesting differences which we will 
highlight as they appear.   
Let us discretize the $(d{+}1)$-dimensional space time and 
put the quantum XY model (\ref{quantum-XY-action}),  
using the usual Villain form, on  
the $(d{+}1)$-dimensional hypercubic lattice: 
\begin{equation}
{\cal S}_{\text{XY}} 
= \frac{1}{2}K_{\text{XY}}  \sum_{j}\sum_{\mu=1}^{d+1} 
\left(\Delta_{\mu}\varphi_{j} - 2\pi m_{j,\mu}\right)^{2} \; .
\end{equation}
The XY coupling $K_{\text{XY}}$ is proportional to $K_{\tau,\perp}$ in 
eq.(\ref{quantum-XY-action}). 
The above XY action should be supplemented by the Berry phase term:
\begin{equation}
{\cal S}_{\text{BP}} = i\, (2S)\sum_{j}{\cal A}_{j,\tau} \; ,
\label{eqn:S-P-BerryPhase}
\end{equation}
where ${\cal A}_{j,\mu}$ ($\mu=1,\ldots,d{+}1$; $x_{d+1}=\tau$), 
the lattice version of the spin gauge field (or equivalently the ${\rm CP}^{1}$ gauge field),   
is half the area of a spherical triangle enclosed by a triad of unit vectors 
$\boln_{0}$ (a fixed reference vector which can be chosen arbitrarily), $\boln(j)$ and $\boln(j+\hat{\mu})$.   
Unlike the corresponding expression for $S=1/2$ antiferromagnets in the 
{\em absence} of magnetic field\cite{Sachdev-Park}, ${\cal
S}_{\text{BP}}$ here does not contain an sign-alternating factor, 
reflecting the canted nature of the classical ground state.  
In addition to these, we add (by hand) the simplest `Maxwellian term' allowed both by 
the arbitrariness of $\boln_{0}$ and by the $2\pi$-redundancy 
(compactness) of ${\cal A}_{j,\mu}$.  
The (1+1)d expression 
(we will be dealing with this case in most of the equations to follow), 
for example, is   
\begin{equation}
{\cal S}^{\text{(1+1)d}}_{\text{Maxwell}} = 
\frac{1}{2 g_{\text{m}}^{2}}
\sum_{j}\left(\epsilon_{\mu\nu}\Delta_{\mu}{\cal A}_{j,\mu}
- 2\pi q_{j^{\ast}}\right)^{2} \; .
\end{equation}
In the above equation, the integer field $q_{j^{\ast}}$ resides  
on the dual lattice site. 
Carrying out the Poisson resummation and Gaussian integration, 
we obtain:
\begin{equation}
{\cal S}^{\text{(1+1)d}}_{\text{Maxwell}} = 
\frac{g_{\text{m}}^{2}}{2}\sum_{j^{\ast}} a_{j^{\ast}}^{2} 
+ i \sum_{j}(\epsilon_{\mu\nu}\Delta_{\mu}{\cal A}_{j,\nu})a_{j^{\ast}} 
\; .
\end{equation}
It is convenient to rewrite the Berry phase term, 
eq.(\ref{eqn:S-P-BerryPhase}), as
\begin{equation}
\begin{split}
{\cal S}_{\text{BP}} &= i\,(2S)\sum_{j}\sum_{\mu}\delta_{\mu,\tau}
{\cal A}_{j,\mu} \\
&= i\,(2S)\sum_{j,\mu}(\epsilon_{\mu\nu}\Delta_{\mu}{\cal A}_{j,\nu})
a_{j^{\ast}}^{(0)} \; ,
\end{split}
\label{eqn:S-P-BerryPhase-2}
\end{equation}
where we have introduced an off-set gauge field $a^{(0)}_{j^{\ast}}$ 
which satisfies 
$\epsilon_{\mu\nu}\Delta_{\nu}a_{j^{\ast}}^{(0)}=\delta_{\mu,\tau}$. 
This can be solved explicitly as, e.g.,  
\begin{equation}
a^{(0)}_{j^{\ast}} = - j^{\ast}_{x} \; .
\end{equation}
As an important physical aside, we note that for the canted spin configuration of eq.(\ref{unit vector}), 
the lattice curl $\epsilon_{\mu\nu}\Delta_{\mu}{\cal A}_{j,\nu}$ (which usually represents the 
antiferromagnetic spin chirality fluctuations) 
is proportional to the vorticity $\epsilon_{\mu\nu}\Delta_{\mu}m_{j,\nu}$ of 
the XY spins on a plaquette surrounding $j^{\ast}$. (This is similar to the situation of the easy plane antiferromagnet 
in the absence of a magnetic field treated in ref.\onlinecite{Sachdev-Park}.)
In terms of the latter, 
the Berry phase of eq.(\ref{eqn:S-P-BerryPhase-2}) reads:
\begin{equation}
{\cal S}_{\text{BP}} = i\,2\pi(S-m)\sum_{j}
(\epsilon_{\mu\nu}\Delta_{\mu}m_{j,\nu})a_{j^{\ast}}^{(0)} \; .
\end{equation}
Notice that if we identify $\epsilon_{\mu\nu}\Delta_{\mu}m_{j,\nu}$ with $q_{j}$ 
of the previous section, 
the above ${\cal S}_{\text{BP}}$ 
exactly coincides with the first line of eq.(\ref{vortex BP factors}).   
This is readily generalized to higher-dimensional cases. 
For example, in (2+1)d we 
may write 
\begin{equation}
{\cal S}_{\text{BP}} = i\, 2\pi(S-m)\sum_{j,\mu}
(\epsilon_{\mu\nu\rho}\Delta_{\nu}m_{j,\rho})a^{(0)}_{j^{\ast},\mu} 
\end{equation}
with 
$(a^{(0)}_{j^{\ast},x},a^{(0)}_{j^{\ast},y},a^{(0)}_{j^{\ast},\tau})
=(-j^{\ast}_{y}/2,j^{\ast}_{x}/2,0)$.  
It is easy to verify that this reproduces the previous result 
(\ref{vortex BP factors}). 

\begin{widetext}
Rescaling $a_{j^{\ast}} \mapsto (2S)a_{j^{\ast}}$ in 
${\cal S}^{\text{(1+1)d}}_{\text{Maxwell}}$ 
and 
collecting the terms 
${\cal S}_{\text{XY}}$, 
${\cal S}^{\text{(1+1)d}}_{\text{Maxwell}}$ 
and ${\cal S}_{\text{BP}}$, we arrive at the following action:
\begin{equation}
\begin{split}
Z^{\text{(1+1)d}} &= \sum_{\{m_{j,\mu}\}}\sum_{\{a_{j^{\ast}}\}}
\int_{0}^{2\pi}\!\prod_{j}d\varphi_{j} 
\exp\biggl[
-\frac{1}{2}K_{\text{XY}}
\sum_{j,\mu}(\Delta_{\mu}\varphi_{j} - 2\pi m_{j,\mu})^{2} \\
& \phantom{\sum_{m_{j,\mu}}\sum_{a_{j^{\ast}}}}
-\frac{g_{\text{m}}^{2}}{2}\sum_{j^{\ast}}
(a_{j^{\ast}} - a_{j^{\ast}}^{(0)})^{2}
- i\,2\pi(S-m)\sum_{j}(\epsilon_{\mu\nu}\Delta_{\mu}m_{j,\nu})a_{j^{\ast}}
\biggr] \; .
\end{split}
\label{eqn:XY-U1-1D}
\end{equation}
\end{widetext}

To make further progress we now need to fix the coefficient of the Berry phase term. 
Here a crucial difference arises with eq.(53) of ref.\onlinecite{Sachdev-Park}, which 
corresponds to setting $m=0$ (and $S=1/2$) in our eq.(\ref{eqn:XY-U1-1D}). By varying $m$, we  
are able to probe through a variety of different effective theories, each characterized by 
a different set of vortex Berry phase factors. Below we will inspect a few representative cases.  

When $S-m \in \mathbb{Z}$, the U(1) gauge field $a_{j^{\ast}}$ and 
the vortices decouple from each other. The $a$-summation in 
eq.(\ref{eqn:XY-U1-1D}) can then be carried out trivially, to yield 
the classical XY model in $(d+1)$ ($d$=1 here) dimensions:
\begin{equation}
\begin{split}
& Z^{\text{(1+1)d}} \\ 
& \sim 
\sum_{\{m_{j,\mu}\}}
\int_{0}^{2\pi}\!\prod_{j}d\varphi_{j} 
\exp\biggl[
-\frac{K_{\text{XY}}}{2}
\sum_{j,\mu}(\Delta_{\mu}\varphi_{j} - 2\pi m_{j,\mu})^{2}
\biggr] \\
&= Z^{\text{2d}}_{\text{XY}} \; .
\end{split}
\end{equation}
Thus in agreement with our arguments of section \ref{sec:vortex-ham}, 
we expect to encounter the usual classical XY transition triggered by 
vortex proliferation as the XY-coupling $K_{\text{XY}}$ is varied.   

A more interesting situation arises for $S-m \in \mathbb{Z}+1/2$.  
Here  the partition function 
(\ref{eqn:XY-U1-1D}) depends on the {\it parity} of the link variable  
$m_{j,\mu}$, 
which leads us to introduce the Ising variable 
$s_{j,j+\hat{\mu}}$ through\cite{Sachdev-Park}:
\begin{equation}
m_{j,\mu} = 2f_{j,\mu} + \frac{1-s_{j,j+\hat{\mu}}}{2} 
\quad (f_{j,\mu}\in \mathbb{Z},s_{j,j+\hat{\mu}}=\pm 1)\; .
\end{equation}
Plugging this into eq.(\ref{eqn:XY-U1-1D}) and carrying out 
the summation over the $a_{j^{\ast}}$ explicitly, 
we obtain\cite{footnote-2} the following partition function:
\begin{equation}
Z^{\text{(1+1)d}} = \sum_{\{s_{j,j+\hat{\mu}}\}}\int_{0}^{2\pi}\!\prod_{j}
d\varphi_{j}\, \be^{-{\cal S}_{\text{Ising}} -{\cal S}_{\text{matter}}} \;
,
\end{equation}
where the action of the quantum $\mathbb{Z}_{2}$ gauge theory 
${\cal S}_{\text{Ising}}$ and its coupling to the matter (XY) field 
each take the form:
\begin{subequations}
\begin{align}
& {\cal S}_{\text{Ising}} = -K_{\text{IGT}}(g_{\text{m}}) 
\sum_{\Box}\left(
\prod_{\Box}s_{j,j+\hat{\mu}} \right) 
+i\pi\, \sum_{j}\frac{1-s_{j,j+\hat{\mu}}}{2} 
\label{eqn:Z2-gauge-action} \\
& {\cal S}_{\text{matter}} = 
-4 K_{\text{XY}} \sum_{j,\mu}s_{j,j+\hat{\mu}}
\cos \left(\frac{1}{2}\Delta_{\mu}\varphi_{j}\right) \; .
\label{eqn:matter-action}
\end{align}
The effect of the Berry phase term of (\ref{eqn:XY-U1-1D}) is 
now encoded 
in the second term of ${\cal S}_{\text{Ising}}$.  
The gauge coupling $K_{\text{IGT}}$, a monotonically decreasing function of $g$, 
is defined through the relation: 
\begin{equation}
\be^{2K_{\text{IGT}}(g)} \equiv 
\sum_{n\in\mathbb{Z}} \be^{-\frac{g^{2}}{2}n^{2}}
/\sum_{n\in\mathbb{Z}} (-1)^{n}\be^{-\frac{g^{2}}{2}n^{2}}
 \; .
\end{equation}
\end{subequations} 
The appearance of the factor 1/2 in the cosine of 
(\ref{eqn:matter-action}) is worth noting; 
it indicates that the Ising gauge field couples to 
{\it fractionalized bosons}, 
whose creation operators can be identified as  
$b_{j}^{\dagger}\sim{\rm e}^{\frac{i}{2}\varphi_{j}}$.
A theory with the same basic structure, i.e. 
fractionalized bosons held together by  
Ising gauge fields, 
is obtained in (2+1) dimensions as well. 
Taking the limit $K_{\text{IGT}}$=0, 
an explicit summation over the gauge variables $s_{j,j+\hat{\mu}}$ 
leaves us with a theory in which coefficients are doubled 
wherever $(\Delta_{\mu}\varphi_{j})$ arises; 
hence we can expect the fractional bosons to be confined in this limit.  
A more nontrivial phase may emerge by looking into regions with 
larger $K_{\text{IGT}}$-values.     
It turns out in fact, that {\it deconfinement} can occur in (2+1) dimensions  
for large $K_{\text{IGT}}$, realizing 
a fractionalized plateau phase 
with topological order.
Such details are most conveniently analyzed by going back to 
the form of eq.(\ref{eqn:XY-U1-1D}) and performing, 
in similarity to the previous section, 
a duality transformation.  

\begin{widetext}
Apart from the difference in the expression for $a_{j^{\ast}}^{(0)}$, 
the derivation of the dual theory 
proceeds according to the prescription of ref.\onlinecite{Sachdev-Park}.  
The final form for (1+1)d reads:  
\begin{equation}
Z^{\text{1d}} =  
\sum_{\{a_{j}^{\ast}\}}\sum_{\{n_{j}^{\ast}\}} \exp \biggl[
-\frac{1}{2}g^{2}_{\text{m}}\sum_{j^{\ast}}
(a_{j^{\ast}}-a_{j^{\ast}}^{(0)})^{2}
-\frac{1}{2K_{\text{XY}}}
\sum_{j,j^{\prime}}\sum_{\mu,\nu}
M_{j,j^{\prime}}^{\mu,\nu}
\epsilon_{\mu\lambda}\Delta_{\lambda}\left\{n_{j^{\ast}}
-(S-m)a_{j^{\ast}}\right\}
\epsilon_{\nu\rho}\Delta_{\rho}
\left\{n_{j^{\ast,\prime}}-(S-m)a_{j^{\ast,\prime}}\right\}
\biggr]  \; ,
\label{eqn:dual-action}
\end{equation}
where $\{n_{j^{\ast}}\}$, whose lattice curl is the boson density, is an integer-valued field 
defined on the dual sites. 
The kernel  
$M_{j,j^{\prime}}^{\mu,\nu}$  
governing the density-density interaction reduces to  
$M_{j,j^{\prime}}^{\mu,\nu}=\delta_{jj^{\prime}}\delta_{\mu\nu}$
for the simplest case of spin systems with short-ranged interactions; 
here it is written in a form which accounts for a more generic Hamiltonian\cite{footnote-K}.   
A similar dual action is also obtained for the (2+1)d case. While we have displayed the 
form of the dual theory for general $S-m$, we continue to concentrate here on the 
case $S-m\in \mathbb{Z}+1/2$.    
\end{widetext}

\begin{figure}[H]
\begin{center}
\includegraphics[scale=0.3]{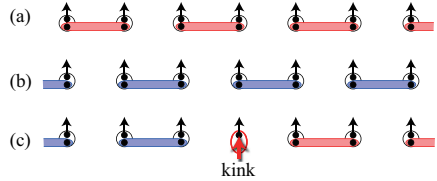}
\end{center}
\caption{
A typical spin state found at an $m=1/2$ plateau in spin-1 chains 
($S-m=1/2$).  Lattice translation symmetry is broken spontaneously 
and as a consequence we have two degenerate ground states (a) and (b). 
Half (arrows) of each spin-1 degree of freedom is frozen by a strong magnetic field 
and the remaining fluctuating half forms valence-bond solid states%
\cite{Nakano}. 
The lowest excitation is a kink which carries an $S^{z}$ quantum number of 
1/2.%
\label{fig:VB-1D}}
\end{figure}
\begin{figure}[ht]
\begin{center}
\includegraphics[scale=0.35]{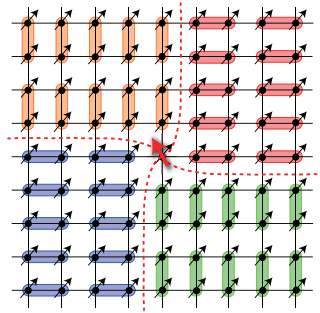}
\end{center}
\caption{
Two-dimensional analogue of the spin-1 partially-polarized valence-bond-solid 
state shown in Fig.\ref{fig:VB-1D}.  Again, the unpolarized `fractions' 
of the spin-1's form a valence-bond pattern. 
Dashed lines denote domain walls separating four different valence-bond patterns. 
At the core of the $\mathbb{Z}_{4}$-vortex is a fractionalized excitation 
carrying $S^{z}=1/2$ (large arrow). The resemblance of the spatial patterns  
as well as the fractionalized nature of the core excitation with those of 
Refs.\onlinecite{Levin,Seradjeh-W-F} is evident. %
\label{fig:VB-2D}}
\end{figure}

Insight into the phase structure of $Z^{\text{1d}}$ 
(and $Z^{\text{2d}}$) comes from 
following the analysis carried out in ref.\onlinecite{Sachdev-Park}.  
We begin by considering the limit $g_{\text{m}}\rightarrow \infty$. 
Under this condition, the fluctuations in the $a_{j^{\ast}}$-field are quenched 
$a_{j^{\ast}}\mapsto a_{j^{\ast}}^{(0)}$ and the model 
(\ref{eqn:dual-action}) reduces (for a diagonal $M_{j,j^{\prime}}^{\mu,\nu}$) 
to the dual vortex model of appendix 
\ref{appendix on duality} (compare with eqs.(\ref{offsetted lagrangian}) and 
(\ref{solution-of-constraint})). 
(Conversely, eq.(\ref{eqn:dual-action}) extends  
the standard dual-vortex model treated in appendix \ref{appendix on duality} 
to allow for fluctuations of the background boson density.) 
The latter is expected to be in a superfluid (or, XY) 
phase\cite{van Otterlo}, 
which persists  
into the region 
$K_{\text{XY}}\gg 1$.  

Turning next to generic values of $g_{\text{m}}$,  
the $a_{j^{\ast}}$-field starts to retain its dynamics and induces fluctuations 
in the background boson density (spin density in the present physical context)  
$\epsilon_{\tau\lambda}\Delta_{\lambda}a_{j^{\ast}}$.      
These fluctuations may stabilize a new  
phase by forming bond-centered orders. 
(By retracing our steps backwards through the sequences described above, 
one sees that the $g_{\text{m}}$-terms of eq.(\ref{eqn:dual-action}) and its 
(2+1)d counterpart are 
 associated with the energies of the spatial links of the direct lattice).   
In fact, by recasting our dual theory into a 2d height model\cite{Sachdev-Park} 
(a 3d frustrated Ising model\cite{Jalabert-S-91} for the (2+1)d case) 
or to a field-theoretical model similar to those studied in ref.\onlinecite{Lannert}, 
it is possible to show that 
for small $K_{\text{XY}}$, {\it valence-bond-solid} states are realized 
in the background of polarized spins moments ($m$ per site), as depicted in 
Figs.\ref{fig:VB-1D} and \ref{fig:VB-2D}  
(for (1+1) and (2+1) dimensions respectively). 
These states support   
kink-like excitations with the fractional quantum number  
$\delta S^{z}=1/2$. 
The 1d version of this valence-bond-solid plateau has been 
observed numerically\cite{Nakano}. It is interesting to note 
the apparent similarity of the $\mathbb{Z}_{4}$-vortex structure 
shown in Fig.\ref{fig:VB-2D} with those studied recently 
by other authors\cite{Levin,Seradjeh-W-F} 
in the context of fractionalized excitations in (2+1)d electron/spin models. 
In addition to these exotic states, 
more conventional `collinear' (i.e. CDW) phases 
may appear when off-diagonal parts of $M^{\mu,\nu}_{i,j}$ 
are sufficiently large. 

\begin{figure}[ht]
\begin{center}
\includegraphics[scale=0.5]{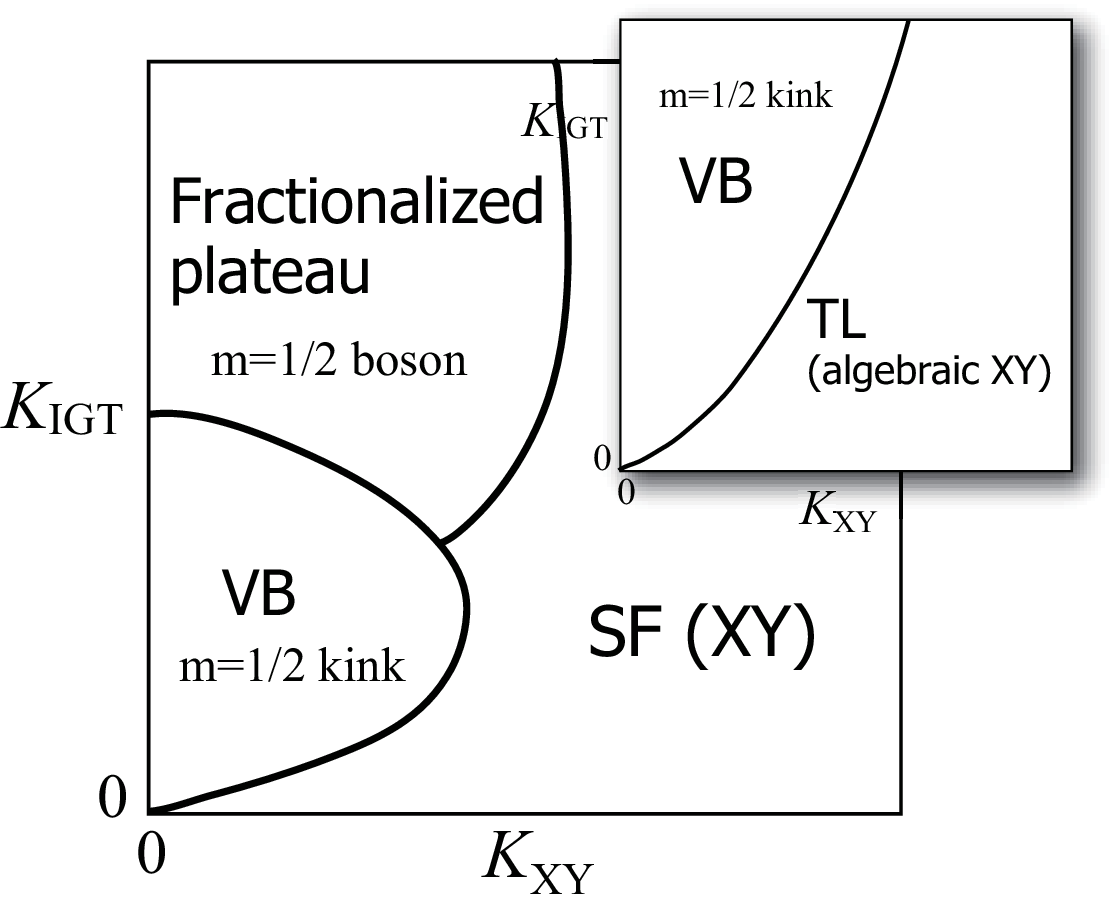}
\end{center}
\caption{
Schematic phase diagram of the fractionalized boson coupled to 
$\mathbb{Z}_{2}$-gauge theory for (1+1)d (inset) and (2+1)d.  
The large-$K_{\text{XY}}$ region is dominated by an XY-ordered 
(or, superfluid(SF)) phase, which, in (1+1)d, is replaced with 
an algebraic Tomonaga-Luttinger(TL) liquid phase.  
In the small-$K_{\text{XY}}$ plateaus region, there is always 
a valence-bond(VB) phase with broken translation symmetry. 
Both in (1+1)d and in (2+1)d, the VB phase supports topological 
excitations (kinks in (1+1)d and $\mathbb{Z}_{4}$ vortices in 
(2+1)d).   
On top of this, there is a {\em featureless} plateau phase with 
fractionalized (i.e. $\delta S^{z}=1/2$) bosons in (2+1)d.   
\label{fig:IGT-phase-diag}}
\end{figure}

Since the 3d frustrated Ising model 
(equivalent to the (2+1)d version of eq.(\ref{eqn:dual-action}) with 
$K_{\text{XY}}=0$) 
has an order-disorder 
transition\cite{Jalabert-S-91}, the phase diagram in the (2+1)d 
case is richer than its (1+1)d counterpart; 
a sufficiently small $g_{\text{m}}$ places the frustrated 
Ising model in a high-temperature paramagnetic phase 
(in this language the ordered phase with broken square-lattice symmetry translates 
into the columnar valence-bond-solid phase) which may be identified with 
the fractionalized phase of the model 
(\ref{eqn:Z2-gauge-action},\ref{eqn:matter-action}) with 
the full space-group symmetry.  
This suggests that {\em non-crystalline} plateau states are 
possible  in $d\geq 2$ 
while in 1d, as implied by the LSM arguments,  
the appearance of plateaus is closely 
tied to the formation of crystalline states. 
(The dual vortex picture affords us with an alternative, simple description of the fractionalized excitations (spinons) 
which characterize this phase: they are the 2$\pi$ vortices of the dual Ginzburg-Landau-type field theory which describes 
the condensate of doubly quantized vortices\cite{Lannert}.
The $\mathbb{Z}_{4}$-kinks in the valence-bond-soid phase, while requiring a subtler analysis, can also be 
treated within a similar field theory language.  
)    
We summarize the above discussion in the form of a phase diagram, depicted schematically in 
Fig.\ref{fig:IGT-phase-diag}.   

In concluding this section, we note  
that the two cases considered above, 
$S-m \in \mathbb{Z}$ and $S-m \in \mathbb{Z}+1/2$, 
each fall into the class of effective theories 
for easy plane antiferromagnets (with no magnetic fields and hence $m=0$)  
with integer and half-integer\cite{Sachdev-Park} $S$. 
As we have already mentioned, however the present problem allows us to  
further extend this approach to other values of $S-m$.     
For instance the argument applies after appropriate modifications 
to the case with $S-m=\mathbb{Z}+1/3$; 
now the original boson is coupled to a $\mathbb{Z}_{3}$ gauge field, 
and is fractionalized in such a way that it carries the `charge' 
$\delta S^{z}=1/3$.  
The detailed analysis for general rational values of $S-m$ will be
reported elsewhere.   

\section{Conclusions}
\label{sec:conclusion}
The message of this article is twofold: 
(1) geometric phases provide a natural language to understand 
how commensurability conditions impose constraints 
on the generation of magnetization in antiferromagnets, 
and (2) the analysis based on this language suggests 
a phase structure which accommodates a wealth of exotic phases.   

In the earlier portions of the   
article we found that a 
geometrical interpretation of the LSM approach emerges 
when we view the spatial modulation of the slow variables 
as an `` adiabatic evolution" along the spin chain. 
Continuing to focus on geometric phases,  
we were subsequently lead  
to effective field theories featuring vortex Berry phase factors  
which are valid in arbitrary dimensions. 
Taken together, these findings afford us with an explanation of  
a curious aspect of the LSM-style momentum-counting argument\cite{OYA,Oshikawa2} 
which was mentioned back in subsection \ref{subsec:LSM}; 
its ability to produce the correct criteria for the 
existence/nonexistence of a magnetization plateau 
despite the seemingly remote connection between the 
plateaus and the momenta of charge-neutral ($\delta S^{z}=0$) 
excitations. The resolution lies in the fact that 
the LSM momentum shift and the action ${\cal S}^{\text{vortex}}_{\text{BP}}$ governing the 
low-energy behavior via topological interference 
stems from  a single quantity--the Berry phase of the canted spins.

The approach exploited in the preceeding two sections 
is intended to serve as the basis for 
seeking as-yet undetected novel phases   
whose existence our theory suggests.    
Indeed the $Z_{2}$ gauge theory of section \ref{sec:Z2-gauge}, 
which is closely linked to the issue of fractionalization en route the 
pairing of vortices\cite{Balents1,Balents2}\, strongly suggests 
that antiferromagnets in an external magnetic field provides a promising playground 
to search for fractionalized spin liquids in spatial dimensions higher than one. 
(A magnetization plateau at $m=1/2$ in translationally invariant $S=1$ Hamiltonians 
on a 2d square lattice, a signature of such a phase, 
may in this regards be an interesting problem to pursue 
numerically as well as experimentally.)    
How the perturbation brought on by spin currents as briefly discussed towards the end of 
section \ref{sec:dual-vortex} affects the magnetization property is another problem that 
we believe warrant further investigations.  
We hope to return to such issues in the future.

%
\begin{acknowledgements}
We thank Y. Hatsugai, T. Hikihara, J. Inoue, M. Kohno, K. Matsushita, 
M. Nakamura, Y. Nonomura, M. Oshikawa and M. Sato 
for helpful discussions. AT acknowledges the hospitality extended by 
K.-S. Kim at KIAS where an early portion of this study was conducted, 
which was later combined with unpublished 
notes of KT. He also thanks A. G. Abanov for pointing to several 
important references on 
phase slips in 1d superconductors.   
Finally we wish to thank the organizers 
of the workshop/symposium 
{\it Topological Aspects of Solid State Physics} 
held at ISSP, Kashiwa and YITP, Kyoto 
where this work was concluded. 
This work was supported in part by 
Grant-in-Aids for Scientific Research (C) 18540386, 
(C) 20540386
(AT) and (C) 18540372, (C) 20540375 (KT) 
from MEXT, Japan. AT and XH are also supported by the 
World Premiere International Research Center Initiative 
of Materials Nanoarchitectonics, MEXT, Japan. 
KT is supported by the global COE program `The next generation of 
physics, spun from universality and emergence' of Kyoto University. 
\end{acknowledgements}
\appendix
\section
{
More on the Berry phase approach to the OYA argument
}
\label{appendix on OYA}
In this appendix we discuss how the 
arguments of section \ref{sec:LSM} are to be extended 
when one wishes to deal with 
a {\it superposition} of spin coherent states 
$\vert\{ {\bf n}(x) \}\rangle$. 
While the case treated in the main text already captures the essential geometric 
properties inherent in the LSM approach, 
this generalization intends to fill in the gap 
between the treatment of the actual ground state of a Heisenberg 
antiferromagnet (which possesses rotational invariance)  
with that given in the main text for a spin coherent state which 
is, in a semiclassical sense polarized in a particular orientation. 

Let $\vert \Psi\rangle$ then be our ground state. We can generally 
expand this state in the (overcomplete) spin coherent states basis $\vert \{ \phi(x) \}\rangle$, 
where in accordance with the main text, 
the low energy physics is described in terms of the angular field 
$\phi(x)$ specifying the in-plane orientation of the staggered spin
component perpendicular to the magnetic field.   
\begin{equation}
\vert \Psi\rangle=\int{\cal D}\phi(x)
\Psi[\phi(x)]\vert\{\phi(x)\}\rangle.  
\label{eqn:decomposition}
\end{equation} 
The wave functional 
$\Psi[\phi(x)]=\langle\{\phi(x)\}\vert\Psi\rangle$ is subject to the 
normalization condition $\int {\cal D}\phi(x)\vert\Psi[\phi(x)]\vert^{2}=1$. 
We will adopt the action ${\cal
S}_{eff}[\phi(\tau,x)]$ given in the text (eq.(\ref{quantum-XY-action})), 
though the arguments to follow apply to a wider 
variety of systems (as long as the 
coefficient of the topological term remains the same). 
The Hamiltonian derived from this action is 
\begin{equation}
{\hat{\cal H}}=
\int dx \left\{ \frac{1}{2K_{\tau}}({\hat \pi}-(S-m))^{2}+\frac{K_{\perp}}{2}
(\partial_{x}\phi)^{2} \right\} \; , 
\end{equation}
where ${\hat \pi}(x)=-i\frac{\delta}{\delta\phi(x)}$ is canonically 
conjugate to $\phi(x)$. 
The appearance of the gauge field-like piece $S-m$ in the kinetic energy term 
has its origin in the topological term.  
In the superfluid analogy, it is simply the
offset value of the superfluid density.  
The functional Schr\"odinger equation reads 
${\hat{\cal H}}({\hat {\pi}}(x), \phi(x))\Psi[\phi(x)]=E\Psi[\phi(x)]$.  

As before we wish to compare the expectation values of the 
crystal momentum in the ground state and the LSM-twisted state. 
We recall that for the pure spin coherent state treated in section \ref{sec:LSM}, 
it was for this purpose essential that the quantity
$\langle\{ {\bf n}(x) \}\vert{\hat T} \vert\{ {\bf n}(x) \}\rangle$
pick a Berry phase factor associated with a round trip experienced 
by the vector ${\bf n}$. 
It turns out however that here the corresponding expression,
$\langle{\Psi}\vert{\hat T}\vert{\Psi}\rangle
=\int {\cal D}\phi(x)\Psi^{\ast}[\phi(x-a)]\Psi[\phi(x)]$
is {\it not} equipped with the anholonomy 
one expects to find when $\phi$ is sent on a round excursion. 
This can be traced to the 
single-valuedness of $\Psi[\phi(x)]$ which implies that the 
wave functional be invariant under a $2\pi$-shift of $\phi(x)$. 
Hence as it stands one cannot straightforwardly    
incorporate the Berry phase effect, 
which we know from the main text to be crucial 
in obtaining the correct expression for the crystal momentum.

The resolution comes from seeking 
an analogy to the quantum mechanical problem of a 
particle on a ring threaded by an Aharonov-Bohm (AB) flux\cite{Altland-Simons} 
to which our problem reduces when $K_{\perp}=0$. 
As in the treatment of the latter, 
the AB-like anholonomy can encoded into the wavefunctional  
through a (singular) gauge transformation
\begin{equation}
{\tilde \Psi}[\phi(x)]=e^{-i\int dx(S-m)\phi(x)}\Psi[\phi(x)],
\end{equation}
which eliminates the AB-gauge field from the Hamiltonian, i.e.  
\begin{equation}
{\tilde {\hat{\cal H}}}=
\int dx \left\{
\frac{1}{2K_{\tau}}{\hat
\pi}^{2}+\frac{K_{\perp}}{2}(\partial_{x}\phi)^{2}
\right\} ,
\end{equation}
at the price of introducing a twisted boundary condition for the 
transformed wavefunctional  
(such that a phase change of $-2\pi(S-m)$ is experienced 
as $\phi(x)$ changes by $2\pi$).

We thus deduce the appropriate generalization of 
$\langle\{ {\bf n}(x) \}\vert{\hat T} \vert\{ {\bf n}(x) \}\rangle$ to be 
\begin{equation}
\begin{split}
& \langle{\tilde \Psi}\vert{\hat T}\vert{\tilde \Psi}\rangle
= \int {\cal D}\phi(x){\tilde \Psi}^{\ast}[\phi(x - a)]
{\tilde \Psi}[\phi(x)]  \\ 
&\qquad = \int{\cal D}\phi(x)e^{-i\int dx(S-m)\partial_{x}\phi}
\Psi^{\ast}[\phi(x - a)]\Psi[\phi(x)],
\label{generalized1}
\end{split}
\end{equation}
where we resorted to continuum notations in the final expression.   
Notice that the phase factor in the second line 
reproduces the correct Berry phase while the 
part $(\Psi[\phi(x+a)])^{*}\Psi[\phi(x)]$ is single-valued.   

Introducing the unitary operator 
${\hat U}_{\rm LSM}=\exp[-i\frac{2\pi}{L}\int dx x{\hat \pi}(x)]$ and 
using the relation 
${\hat U}_{\rm LSM}^{\dagger}{\hat T}{\hat U}_{\rm LSM}
=\exp[-i\frac{2\pi}{L}\int dx{\hat \pi}(x)]{\hat T}$, 
it is easy to see that the 
counterpart of eq.({\ref{generalized1}}) for the LSM-twisted state reads 
\begin{equation}
\begin{split}
& \langle{\tilde \Psi}\vert
{\hat U}_{\rm LSM}^{\dagger}{\hat T}{\hat U}_{\rm LSM}
\vert{\tilde \Psi}\rangle \\
& \qquad =
\int{\cal D}\phi(x)
{\tilde \Psi}^{\ast}\left[
\phi(x-a)-\frac{2\pi}{L}a\right]{\tilde \Psi}[\phi(x)] \\
& \qquad =
\int{\cal D}\phi(x)
\Psi^{\ast}\left[\phi(x-a)-\frac{2\pi}{L}a \right]\Psi[\phi(x)] \\
& \qquad {\mbox{\hspace*{15mm}}}
\times 
e^{i2\pi(S-m)}e^{-i\int dx (S-m)\partial_{x}\phi}
\end{split}
\end{equation}
Due to its single-valued nature, $\Psi[\phi(x)]$ basically factorizes  
into factors of the form $\sum_{n(x)\in{\mathbb{Z}}}C_{n(x)}e^{in(x)\phi(x)}$ 
coming from each point $x$. 
The effect of shifting the angular field 
$\phi(x)$ by $\frac{2\pi}{L}a$ at each of the $L/a$ sites thus 
merely amounts to a net factor of unity, 
i.e. $\Psi[\phi(x)+\frac{2\pi}{L}a]=\Psi[\phi(x)]$. Hence we have 
\begin{equation}
\begin{split}
& \langle{\tilde \Psi}\vert
{\hat U}_{\rm LSM}^{\dagger}{\hat T}{\hat U}_{\rm LSM}
\vert{\tilde \Psi}\rangle \\
& \qquad =
\int{\cal D}\phi(x)
\Psi^{\ast}[\phi(x-a)]\Psi[\phi(x)] \\
& \qquad {\mbox{\hspace*{15mm}}}
\times 
e^{i2\pi(S-m)}e^{-i\int dx (S-m)\partial_{x}\phi}
\label{generalized2}
\end{split}
\end{equation}
A comparison of eqs.(\ref{generalized1}) and (\ref{generalized2}) yields the 
desired quantization rule. 


\section{Boson-Vortex duality in the presence of a topological term}
\label{appendix on duality}
Here we provide the main steps leadings to the Berry phases of 
eq.(\ref{vortex BP factors}), which makes extensive 
use of the boson-vortex duality transformation (performed at the first quantization level). 
Though the technique is standard,  
we feel that it may be worthwhile to illustrate how the presence of a topological term 
brings about modifications and gives rise to vortex Berry phases which ultimately 
dominate the physics at low energy. 
We also sketch how to retain this Berry phase effect when 
one switches from the first to the second quantized description of the vortices.   
\subsection{First quantization}
For simplicity we work in the continuum formulation while being careful to 
take proper account of the Berry phases;   
carrying out a lattice duality transformation  
will of course produce the same results. 
We begin by introducing a Hubbard-Stratonovich auxiliary vector field 
$J_{\mu}=(J_{\tau},{\bf J})$ and rewrite the imaginary-time 
lagrangian density of the effective theory 
${\cal S}[\phi(\tau,{\bf r})]=\int d\tau d^d {\bf r}{\cal L}$ as 
\begin{equation}
\begin{split}
{\cal L} =& 
i\left( J_{\tau}+\frac{S-m}{a^d}\right)\partial_{\tau}\phi
+\frac{J_{\tau}^{2}}{2K_{\tau}}  \\
& +i{\bf J}{\cdot}\nabla\phi+\frac{{\bf J}^{2}}{2K_{\perp}}.
\end{split}
\end{equation}
This is followed by a decomposition of the phase field $\phi$ into components 
with and without vorticity, $\phi=\phi_{\rm v}+\phi_{r}$, where 
$(\partial_{\mu}\partial_{\nu}-\partial_{\mu}\partial_{\nu})\phi_{\rm v}\neq 0$. 
The vorticity-free component $\phi_{r}$ can be safely integrated out, 
which yields the constraint 
\begin{equation}
\partial_{\mu}{\widetilde J}_{\mu}=0,
\label{constraint}
\end{equation}
where ${\widetilde J}_{\mu}=J_{\mu}+\delta_{\mu\tau}\frac{S-m}{a^d}$, 
and the lagrangian density becomes  
\begin{equation}
{\cal L}=i{\widetilde J}_{\mu}\partial_{\mu}\phi_{\rm v}
+\frac{({\widetilde J}_{\tau}-\frac{S-m}{a^d})^{2}}{2K_{\tau}}
+\frac{{\widetilde {\bf J}}^{2}}{2K_{\perp}}.
\label{offsetted lagrangian}
\end{equation}
The divergence-free condition, eq.(\ref{constraint}), is solved explicitly in terms of 
a new field whose form depends on the dimensionality:
\begin{equation}
{\widetilde J}_{\mu}=\left\{
\begin{array}{lr}
\epsilon_{\mu\nu}\partial_{\nu}\varphi & (d=1) \\ 
\epsilon_{\mu\nu\lambda}\partial_{\nu}b_{\lambda} & (d=2) \\
\epsilon_{\mu\nu\lambda\rho}\partial_{\nu}b_{\lambda\rho} & (d=3)
\end{array}
\right.
\label{solution-of-constraint}
\end{equation}
where $\varphi$, $b_{\mu}$, and $b_{\mu\nu}(=-b_{\nu\mu})$ are all 
vorticity-free. Here on the procedures will be described separately for 
each dimensionality $d$.     

\noindent
(i) $d$=1: We define the density of spacetime vortices $\rho^{\rm v}$ via  
$(\partial_{\tau}\partial_{x}-\partial_{x}\partial_{\tau})\phi_{\rm v}
=\sum_{j}2\pi q^{\rm v}_{j}\delta(\tau-\tau^{\rm v}_{j})
\delta(x-x^{\rm v}_{j})\equiv 2\pi\rho^{\rm v}$, 
where $(\tau^{\rm v}_{j},x^{\rm v}_{j})$ is the spacetime coordinate 
of the $j$-th vortex event 
(the notation $X_{j}$ which appeared in eq.(\ref{vortex BP factors}) 
in the main text corresponds here to  
$x^{\rm v}_{j}$.
).
After an integration by parts and a shifting of the 
$\varphi$-field, $\varphi={\widetilde \varphi}+\frac{S-m}{a}x$, we arrive at 
\begin{eqnarray}
{\cal L}
&=&
\frac{1}{2K_{\perp}}(\partial_{\tau}{\widetilde \varphi})^{2}
+\frac{1}{2K_{\tau}}(\partial_{x}{\widetilde \varphi})^{2}
\nonumber\\
&&
+i2\pi\rho_{\rm v}\left({\widetilde \varphi}+\frac{S-m}{a}x\right).
\end{eqnarray} 
Integrating over ${\widetilde \varphi}$ we obtain 
\begin{equation}
{\cal L}=\pi^{2}\rho^{\rm v}\frac{1}{-\partial^{2}}\rho^{\rm v}
+i2\pi\rho^{\rm v}\frac{S-m}{a}x.
\end{equation} 
The first term on the right hand side is the intervortex Coulombic interaction 
for which the kernel $1/(-\partial^{2})$ was defined in the main text. 
The second term represents the vortex Berry phase terms,  
i.e. $i2\pi\int d\tau dx\rho^{\rm v}\frac{S-m}{a}x=i2\pi\sum_{i}q^{\rm v}_{i}\frac{S-m}{a}x^{\rm v}_{i}$. 

\noindent
(ii) $d$=2: We start by establishing some notations. 
The 3-vector $b_{\mu}$ in eq.(\ref{solution-of-constraint}) 
is often referred to as the vortex gauge field or the dual gauge field 
(note the invariance of eq.(\ref{solution-of-constraint}) (in (2+1)d)  
with respect to the gauge transformation $b_{\mu}\rightarrow b_{\mu}+\partial_{\mu}\chi$). 
Dual electric/magnetic fields can be  constructed from these gauge
fields: 
\begin{equation}
\begin{split}
& {\bf E}^{\text{dual}}=(E^{\text{dual}}_{x}, E^{\text{dual}}_{y})
=(\partial_{\tau}b_{x}-\partial_{x}b_{\tau}, 
\partial_{\tau}b_{y}-\partial_{y}b_{\tau}), \\ 
& B^{\text{dual}}=\partial_{x}b_{y}-\partial_{y}b_{x} \; .
\end{split}
\end{equation} 
They are related to the auxiliary field ${\widetilde J}_{\mu}$ through 
\begin{equation}
\begin{split}
&{\widetilde J}_{\tau}=B^{\text{dual}}, \\
&{\widetilde J}_{x}=-E^{\text{dual}}_{y} \; , \;\;
{\widetilde J}_{y}=E^{\text{dual}}_{x} \; .
\end{split}
\end{equation} 
\begin{widetext}
Upon introducing the vortex 3-current 
$J^{\rm v}_{\mu}\equiv
\frac{1}{2\pi}\epsilon_{\mu\nu\lambda}
\partial_{\nu}\partial_{\lambda}\phi_{\rm v}$  
which couples to the dual gauge field, 
the lagrangian density can be rewritten as 
\begin{equation}
{\cal L} =
\frac{1}{2K_{\tau}}\left( B^{\text{dual}}-\frac{S-m}{a^2}\right)^{2} 
+\frac{1}{2K_{\perp}}({\bf E}^{\text{dual}})^{2}
+i2\pi b_{\mu}J^{\rm v}_{\mu}.
\label{append-2d-lagrange1}
\end{equation}
We then divide $b_{\mu}$ (choosing a suitable gauge) 
into two parts, viz.  
$b_{\mu}={\bar b}_{\mu}+\delta b_{\mu}$, where  
${\bar b}_{\mu}$ generates a 
background magnetic field $\partial_{x}{\bar b}_{y}-\partial_{y}{\bar b}_{x}
=\frac{S-m}{a^{2}}$ while $\delta b_{\mu}$ represents the 
deviation from this offset value, 
$\delta B^{\text{dual}} \equiv B^{\text{dual}}-\frac{S-m}{a^{2}}
=\partial_{x}\delta b_{y}-\partial_{y}\delta b_{x}$. 
The gauge is so chosen that ${\bar b}_{\mu}$ 
does not modify the dual electric field ${\bf E}^{\text{dual}}$ (a simple choice being  
$({\bar b}_{\tau},{\bar b}_{x},{\bar b}_{y})=(0,0,\frac{S-m}{a^{2}}x)$). 
Having set up the necessary notations, we can recast 
eq.(\ref{append-2d-lagrange1}) as 
\begin{equation}
{\cal L}
=\frac{1}{2K_{\tau}}(\delta B^{\text{dual}})^{2}
+\frac{1}{2K_{\perp}}({\bf E}^{\text{dual}})^{2} 
+i2\pi\delta b_{\mu}J^{\rm v}_{\mu}+i2\pi{\bar b}_{\mu}J^{\rm v}_{\mu}.
\label{append-2d-lagrange2}
\end{equation}
\end{widetext}
On integrating over $\delta b_{\mu}$ we are left with 
\begin{equation}
{\cal L}={\cal L}_{\text{Coulomb}}
+i2\pi{\bar b}_{\mu}J^{\rm v}_{\mu},
\end{equation} 
where the first term on the right hand side is again the Coulombic interaction 
(now between segments of the spacetime vortex loops) 
which is of the form 
$\pi^{2}J^{\rm v}_{\mu}(\frac{1}{-\partial^{2}})
(\delta_{\mu\nu}-\frac{\partial_{\mu}\partial_{\nu}}{\partial^{2}})
J^{\rm v}_{\nu}$. 
The second term assigns a Berry phase to each spacetime vortex loop, as we will now see. 
Since the vortex current field $J^{\rm v}_{\mu}$ is nonvanishing 
only on such loops (with a magnitude equal to the vorticity $q^{\rm v}_{i}$), we can convert 
the spacetime integration of this term into a sum of contour integrals 
along each vortex loop (we use below a short-hand notation $\vec{V}$ to denote 
a $(d{+}1)$-vector $V_{\mu}$)  :
\begin{equation}
\begin{split}
{\cal S}^{\rm vortex}_{\rm BP}&= 
i\, 2\pi\int d\tau d^{2}{\rm r}\,{\bar b}_{\mu}J^{\rm v}_{\mu}  \\
&=
i\, 2\pi\sum_{j}q^{\rm v}_{j}\oint_{C_{j}}ds \frac{dx^{\rm v}_{j, \mu}(s)}{ds}
{\bar b}_{\mu}(\vec{x}^{\rm v}_{j}(s)).
\label{conversion}
\end{split}
\end{equation}
The parameter $s(\in[0,1])$ is used to specify the distance along each loop contour $C_{j}$. 
Using Stokes' theorem this contour integral can in turn be expressed as an area integral. 
In vectorial notations  
$
i2\pi\sum_{j}q^{\rm v}_{j}\oint_{C_{j}}d{\vec x}^{\rm v}_{ j}
\cdot {\vec {\bar b}}({\vec x}^{\rm v}_{ j})
=i2\pi\sum_{j}q^{\rm v}_{j}\int_{D_{j}} d{\vec A}\cdot {\rm rot}{\vec {\bar b}}
$ where $\partial D_{j}=C_{j}$ and $d{\vec A}$ is the area element on the two-dimensional 
domain $D_{j}$. Recalling that by definition 
${\rm rot}{\vec {\bar b}}=\frac{S-m}{a^2}{\hat e}_{\tau}$, 
the final expression for the Berry phase term is 
$i2\pi(S-m)\sum_{j}q^{\rm v}_{j}A^{xy}_{j}$, where $A^{xy}_{j}$ is the area bounded by the 
xy-plane projection of the loop $C_{j}$. 

 \noindent
 (iii) $d=3$: The procedures involved are natural extensions of 
the $d=2$ case.  
(Readers seeking more details on the framework specific to $d$=3 can consult 
the work of Zee\cite{Zee}.)    
\begin{widetext}
The lagrangian density corresponding to eq.(\ref{append-2d-lagrange1}) is 
\begin{equation}
{\cal L}=\frac{1}{2K_{\tau}}\left(
\epsilon_{ijk}\partial_{i}b_{jk}-\frac{S-m}{a^{3}}
\right)^{2}
+\frac{1}{2K_{\perp}}\left[
(\epsilon_{x\alpha\beta\gamma}\partial_{\alpha}b_{\beta\gamma})^{2}
+
(\epsilon_{y\alpha\beta\gamma}\partial_{\alpha}b_{\beta\gamma})^{2}
\right]
+i\, 2\pi b_{\mu\nu}J^{\rm v}_{\mu\nu},
\end{equation}
\end{widetext}
where the suffixes in roman letters stand for spatial indices, 
and information on the vortex world sheets is encoded in the 
2-form 
${J^{\rm v}_{\mu\nu}}
\equiv\frac{1}{2\pi}\epsilon_{\mu\nu\lambda\rho}\partial_{\lambda}\partial_{\rho}\phi_{\rm v}$. 
Within a worldsheet parametrization scheme using a set of two parameters $(s,u)$, 
the latter can also be written as 
$J^{\rm v}_{\mu\nu}=\sum_{j}q^{\rm v}_{j}\int ds \int du \epsilon^{ab}\partial_{a}
x^{\rm v}_{j,\mu}\partial_{b}
x^{\rm v}_{j,\nu}\delta^{(4)}(\vec{x}-\vec{x}^{\rm v}_{j}(s,u))$, 
where $a$, $b$ are $s$ or $u$ 
(compare with expression for vortex current employed in eq.(\ref{conversion})). 
Since each worldsheet is nothing but the spacetime trajectory of a vortex loop 
living in 3d space, $J^{\rm v}_{\mu\nu}$ can be regarded as a vortex loop current. 
As in the 2+1d case we make the decomposition $b_{\mu\nu}={\bar b}_{\mu\nu}
+\delta b_{\mu\nu}$ with the background field satisfying\ 
\begin{equation}
\epsilon_{ijk}\partial_{i}{\bar b}_{jk}=\frac{S-m}{a^{3}}, 
\label{3d-background-flux}
\end{equation}
together with the condition 
that $\epsilon_{\mu\nu\lambda}\partial_{\mu}{\bar b}_{\nu\lambda}=0$ when 
one of the indices is $\tau$. Integrating out $\delta b_{\mu\nu}$, we get 
\begin{equation}
{\cal L}={\cal L}_{\text{Coulomb}}+i\,2\pi 
J^{\rm v}_{\mu\nu}{\bar b}_{\mu\nu}. 
\end{equation}
The first term is the current-current interaction between worldsheets, 
while the second is the Berry phase term, which contributes 
to the action a sum of surface integrals on each worldsheet.  
The latter converts, upon an application of the Stokes-Gauss theorem 
into 3-volume integrals of the flux 
$\epsilon_{\mu\nu\lambda\rho}\partial_{\nu} {\bar b}_{\lambda\rho}$. 
Since the only nonzero flux component comes from 
eq.(\ref{3d-background-flux}), we find that the vortex Berry phase 
amounts to $i2\pi(S-m)\sum_{j}q^{\rm v}_{j}V^{xyz}_{j}$, 
where $V^{xyz}_{j}$ is the 3-volume of the interior of 
the $j$-th worldsheet projected onto the $xyz$-subspace.   
\subsection{Second quantization (vortex field theory)} 
The dual theories derived above can be used to obtain the vortex field 
theories of eq.(\ref{vortex-field-theory}).  
We give a shortcut version of this procedure for the case of 
$d=2$ which, as mentioned in the main text, 
is interesting in view of its relation to the work of 
Lee and Shankar\cite{Lee-Shankar} (the steps for the cases 
$d=1$ and $d=3$ are essentially the same). 
We first make the following relabelling in eq.(\ref{append-2d-lagrange2}):  
$\delta b_{\mu}\rightarrow{\widetilde b}_{\mu}$, 
$\delta B^{\text{dual}} \rightarrow{\widetilde B}$, and 
${\bf E}^{\text{dual}}\rightarrow{\widetilde {\bf E}}$.  
The action is then decomposed into two parts, 
${\cal S}={\cal S}_{\text{kin}}+{\cal S}_{j\text{-}b}$, 
where 
${\cal S}_{\text{kin}}=\int d^{3}x [\frac{1}{2K_{\tau}}{\widetilde B}^{2}
+\frac{1}{K_{\perp}}{\widetilde {\bf E}}^{2}]$  
is the Maxwellian kinetic energy term, and 
${\cal S}_{j\text{-}b}$ gives the coupling between the vortex current 
and the dual gauge field. 
We write the latter as 
\begin{equation}
{\cal S}_{j\text{-}b}=-i\, 2\pi\int d^{3}xJ^{\rm v}_{\mu}
(\partial_{\mu}{\widetilde \varphi}
-{\widetilde b}_{\mu}-\delta_{\mu y}\rho x),
\end{equation}
where as before the ``superfluid density'' in (2+1)d is 
$\rho=\frac{S-m}{a^2}$, and we have introduced a new scalar field 
${\widetilde \varphi}$ whose role is to impose the continuity 
of the vortex current\cite{Igor}, $\partial_{\mu}J^{\rm v}_{\mu}=0$. 
We now place our system on a spacetime lattice. 
In so doing, note that the vortex current can be written 
(as in eq.(\ref{conversion})) as  
$J^{\rm v}_{\mu}=\sum_{j}q^{\rm v}_{j}\oint_{C_{j}}ds
\frac{dx^{\rm v}_{j\mu}}{ds}
\delta^{(3)}(\vec{x}-\vec{x}^{\rm v}_{j}(s))$. 
In the discretized lattice version 
of the theory, each segment of the vortex current can be represented 
by an integer-valued 3-vector 
$t_{j,\mu}=q^{\rm v}_{j}\frac{dx^{\rm v}_{j,\mu}}{ds}$ defined 
on the dual lattice.   
\begin{widetext}
The partition function of a grand-canonical ensemble of spacetime 
vortex loops is thus 
\begin{equation} 
Z=\int(\prod_{n,\mu} d{\widetilde \varphi}_{n}d{\widetilde b}_{n\mu})
\sum_{\{t_{n\mu}\}}
\prod_{n,\mu}
e^{-{\cal S}_{kin}[{\widetilde b}_{n\mu}]}
e^{-i2\pi t_{n\mu}(\Delta_{\mu}
{\widetilde \varphi}_{n}-{\widetilde b}_{n\mu}-\delta_{\mu}\rho n_{x})}
e^{-\lambda t_{n\mu}^{2}},
\label{grand canonical}
\end{equation}
where $n=(n_{\tau},n_{x},n_{y})$ is the site index, and $\lambda$ 
the line tension of each vortex loop. Using the Poission resummation 
formula, we can trade the sum over $t_{n\mu}$ with an integration 
over a continuous variable $X_{n\mu}$, at the price of introducing 
another set of integers $\{m_{n\mu}\}$.  
After the $X_{n\mu}$-integration is carried out, we obtain   
\begin{equation}
Z=\int(\prod_{n\mu}d{\widetilde \varphi}_{n\mu}d{\widetilde b}_{n\mu})
\sum_{\{m_{n\mu}\}}
\prod_{n\mu}
e^{-S_{kin}[{\widetilde b}_{n\mu}]}
e^{-\left[
\frac{1}{2\lambda}
2\pi(\Delta_{\mu}{\widetilde \varphi}-{\widetilde b}_{n\mu}
-\delta_{\mu y}\rho n_{x}+m_{n\mu}) \right]^{2}},
\end{equation}
which is just the Villain form of 
\begin{equation}
Z=\int(\prod_{n\mu}d{\widetilde \varphi}_{n\mu}d{\widetilde b}_{n\mu})
\prod_{n\mu}
e^{-S_{kin}[{\widetilde b}_{n\mu}]}
e^{-\frac{1}{2\lambda}\cos\left[
2\pi(\Delta_{\mu}{\widetilde \varphi}-{\tilde b}_{\mu}
-\delta_{\mu y}\rho {n_{x}})
\right]}.
\end{equation}
\end{widetext}


\begin{thebibliography}{9}
\bibitem{footnote-2}
The only difference from the treatment in ref.\onlinecite{Sachdev-Park} 
is the expression of $a^{(0)}_{j^{\ast}}$.  After summing over the dual 
sites $j^{\ast}$, we arrive at exactly the same expression as eq.(55) 
of ref.\onlinecite{Sachdev-Park}.
\bibitem{footnote-K}
For instance the introducion of longer-range $S^{z}S^{z}$ interactions will 
generate off-diagonal ($i{\neq} j$) elements of  
$M^{\mu,\nu}_{i,j}$.
\end{thebibliography}
\end{document}